\documentclass[11pt]{article}
\pdfoutput=1
\usepackage{jheppub} 

\usepackage[T1]{fontenc} 
\usepackage{braket}
\usepackage{mathrsfs}
\usepackage{latexsym}
\usepackage{wrapfig}
\usepackage{tabularx}
\usepackage[normalem]{ulem}
\usepackage{feynmp-auto}
\usepackage{mathtools}
\usepackage{tikz}
\usetikzlibrary{decorations.pathreplacing}
\newcommand{\der}{\partial}
\newcommand{\no}{\nonumber}

\newcommand{\disc}{\text{Disc}}
\newcommand{\Mpl}{M_{\rm pl}}
\newcommand{\Ms}{M_{\rm s}}
\newcommand{\Mstar}{M_*}
\newcommand{\mth}{m_\text{th}}
\newcommand{\Lag}{\mathcal{L}}
\newcommand{\scat}{\mathcal{M}}

\newcommand{\tree}{\text{tree}}
\newcommand{\VS}{\text{VS}}
\newcommand{\oneloop}{\text{one-loop}}

\title{
String loops and gravitational positivity bounds: imprint of light particles at high energies
}

\author[a]{Simon Caron-Huot,}
\author[b]{Junsei Tokuda}

\affiliation[a]{Department of Physics, McGill University, 3600 Rue University, Montr\'{e}al, H3A 2T8, QC Canada}
\affiliation[b]{Particle Theory and Cosmology Group, Center for Theoretical Physics of the Universe, Institute for Basic Science (IBS), Daejeon, 34126, Korea}

\emailAdd{schuot@physics.mcgill.ca}
\emailAdd{jtokuda@ibs.re.kr}

\abstract{ 
We study loop corrections to positivity bounds on effective field theories in the context of $2\to 2$ scattering in gravitational theories,
in the presence of light particles.  
It has been observed that certain negative contributions at low energies
are enhanced by inverse powers of a small mass $m$ and are nontrivial to cancel against other low-energy contributions.
These originate from near the forward limit of diagrams involving graviton exchange.
We observe that scattering in this kinematics domain remains infrared-sensitive even at high center-of-mass energy. By considering a string-inspired model in which high-energy loops can be calculated using unitarity and Regge behavior of tree amplitudes, we uncover a natural mechanism
through which $1/m$-enhanced terms perfectly cancel between low and high energy contributions.
This concretely explains possible positivity violations in the presence of gravity from the high-energy viewpoint.
}

\begin{document}
{\baselineskip0pt
\rightline{\baselineskip16pt\rm\vbox to-20pt{
           \hbox{CTPU-PTC-24-17}
\vss}}%
}

\maketitle
\flushbottom

\section{Introduction}
When a quantum field theory involves a tower of higher-spin states, $2\to 2$ scattering amplitudes $\scat_{2 \to 2}(s,t)$ often exhibits a characteristic power-law behavior called Regge behavior,
\begin{align}
    \scat_{2 \to 2}(s,t)
    \propto f(t) s^{\alpha(t)}
    \,,\label{regge_intro}
\end{align}
in some energy range with $s\gg M_\text{spin}^2, |t|$. Here, $s$ and $t$ are the Mandelstam variables and $M_\text{spin}$ denotes a typical mass scale of the tower.  
The amplitude \eqref{regge_intro} is characterized by two functions: $\alpha(t)$ and $f(t)$. The former is the (leading) Regge trajectory which encodes the spectrum of the higher-spin tower. The latter corresponds to the so-called Regge residue which determines how strongly the scattered particles couple to the Regge trajectory. 
Regge behavior has been found in various scattering processes in QCD and string theory.

The basic question we would like to address in this study concerns the sensitivity of $f(t)$ and $\alpha(t)$ at small $|t|$ on infrared (IR) physics.
Even though Regge behavior is only observed in high-energy processes, one cannot assume that these quantities are determined entirely by short-distance physics. How can we characterize IR effects in the range $|t|\ll M_{\rm spin}^2\ll s$?

In the context of QCD, the notion that small-$|t|$ Regge physics is IR sensitive is a familiar one.
Most basically,
$\sqrt{-t}$ in this limit is Fourier-conjugate to impact parameter, so in proton-proton scattering one expects and indeed observe a very rich structure at
the proton radius and QCD scales (see for example ~\cite{ATLAS:2014vxr,TOTEM:2015oop}).
Non-perturbatively, it has long been suggested \cite{Anselm:1972ir,Cohen-Tannoudji:1972gqd,Tan:1974gd,Khoze:2000wk,Jenkovszky:2017efs} that pion-loop corrections would be crucial to explain the observed $t$-dependence of differential cross section in the domain $|t|\lesssim 0.2 \text{ GeV}^2$. Here, the pions play the role of light particles which are much lighter than the typical Reggeization scale $\sim \mathcal{O}(1) \text{GeV}$. In particular, loop corrections to the Pomeron trajectory $\alpha(t)$ have been evaluated using elastic unitarity of $t$-channel partial waves, which predicts the presence of a two-pion threshold singularity at $t=4m_\pi^2$~\cite{Anselm:1972ir}. In the perturbative domain, where Regge behavior can be calculated from a composite state of gluons known as the BFKL Pomeron \cite{Kuraev:1977fs,Balitsky:1978ic},
it is also known that the strong coupling $\alpha_s(|t|)$ runs with $|t|$ rather than with $s$ \cite{Fadin:1998py}. As a last example, the calculable heavy quark contribution to the coupling between the BFKL Pomeron and two photons is known to be singular at $t=4M_{\rm quark}^2$, even when $s$ is much larger \cite{Nikolaev:1990ja,Kovchegov:1999kx}. All of these indicate that the quantities $\alpha(t)$ and $f(t)$ must be assumed to be sensitive to physics at the scale $t$.

In this paper we will be interested in a different context, in which the ultraviolet (UV) completion of an effective field theory (EFT) of gravity displays Reggeization  \eqref{regge_intro} of the graviton in some energy range $M_{\rm spin}^2\lesssim s \ll M_{\rm Planck}^2$.
This is certainly the case in weakly coupled string theory, and this is indeed where our controlled examples will be taken from; however our discussion will be quite general.  In this scenario,  Reggeization is associated with heavy string excitation of mass $\geq \Ms$, where $\Ms$ denotes the string scale; hence, we have $\Ms=M_\text{spin}$.  We would like to determine whether string loops, in the presence of light particles of mass $m\ll \Ms$, can produce terms enhanced by $1/m$ in the functions $\alpha(t)$ or $f(t)$ or their derivatives.

One motivation for this question comes from positivity sum rules, which are used to constrain the space of low-energy 
EFTs that admit an
UV completion compatible with causality and unitarity.
Gravity complicates such sum rules due to the presence of a ``graviton pole'' which precludes taking
forward limits $t=0$. Nonetheless, assuming
Regge behavior~\cite{Tokuda:2020mlf} in some energy range, sum rules for low-energy coefficients have been derived which take the following schematic form
(see also~\cite{Hamada:2018dde,Herrero-Valea:2020wxz, Alberte:2021dnj,Herrero-Valea:2022lfd,Noumi:2022wwf} and section~\ref{sec:review} below for more details):
\begin{align}
    c_2(0)
    \sim
    \text{(positive term)}
    +
    \frac{1}{\Mpl^2}
    \left[
        -\der_t(\ln f(t))
        +
        \der_t\ln (\der_t\alpha(t))
    \right]_{t=0}
    \,.\label{eq:sumrule_intro}
\end{align}
The first term on the right-hand side (RHS) is not calculable within a given EFT but its positivity is ensured by unitarity. The second term, the sum of terms in the square brackets, is sign indefinite. This makes it non-trivial to decide whether the non-negativity condition $c_2(0)\geq 0$ is valid. By contrast, in the absence of gravity, the sum rule becomes $c_2(0)\sim  \text{(positive term)} >0$, giving the so-called positivity bounds~\cite{Pham:1985cr,Adams:2006sv}. 

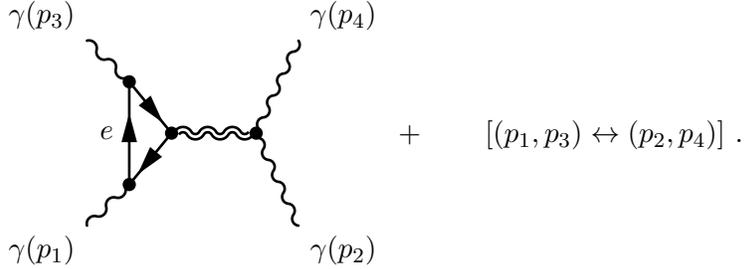
\begin{figure}[t]
\centering
\begin{fmffile}{photon-photon}
\begin{align*}
\raisebox{-0.45\height}{
\begin{fmfgraph*}(100,70)
\fmfleft{i1,o3}
\fmfright{i2,o4}
\fmf{photon}{v3,o3}
\fmf{photon}{i1,v1}
\fmf{photon}{v2,o4}
\fmf{photon}{i2,v2}
\fmf{fermion,label=$e$,label.side=left,tension=-0.1}{v1,v3}
\fmf{fermion}{v4,v1} 
\fmf{dbl_wiggly}{v2,v4}
\fmf{fermion}{v3,v4}
\fmfdot{v1,v3,v2,v4}
\fmflabel{$\gamma(p_1)$}{i1}
\fmflabel{$\gamma(p_2)$}{i2}
\fmflabel{$\gamma(p_3)$}{o3}
\fmflabel{$\gamma(p_4)$}{o4}
\end{fmfgraph*}
}
\qquad
+
\qquad
\left[(p_1,p_3)\leftrightarrow (p_2,p_4)\right]
\,.
\end{align*}
\end{fmffile}
\caption{$\gamma\gamma\to \gamma\gamma$ scattering mediated by the $t$-channel graviton exchange at one-loop level. A solid line with an arrow, a single wiggly line, and a double wiggly line represent an electron, photon, and a graviton, respectively.}
\label{fig:ph_grav1}
\end{figure}

The implications of \eqref{eq:sumrule_intro} become more precise if the $t$-dependence of $f(t)$ and $\alpha(t)$ can be evaluated. In fact, what is most desirable from a phenomenological perspective is to precisely evaluate their dependence on
light mass scales $m$. This is because, for a given EFT, calculable loop effects below the cutoff typically produce large negative terms in $c_2(0)$ enhanced by $m^{-2}$: e.g., in the case of $2\to2$ scattering of photons in the Standard Model, 
the one-loop diagrams shown in fig.~\ref{fig:ph_grav1} give a negative term as~\cite{Cheung:2014ega}\footnote{Strictly speaking, a sum over helicities is
taken to construct a $s\leftrightarrow u$ symmetric amplitude~\cite{Cheung:2014ega}.}
\begin{align}
    c_2(0)|_\text{fig.\ref{fig:ph_grav1}}
    =
    -\frac{11}{90\pi^2}\frac{e^2}{\Mpl^2m_e^2}
    \,.\label{c2_eg1_intro}
\end{align}
Here, $e$ and $m_e$ denote the electric charge and the electron mass, respectively. If this negative term could not be explained by the square bracket in \eqref{eq:sumrule_intro}, we would have
to conclude that additional low-energy physics is required to give positive terms on the left-hand side (LHS) which are enhanced by at least $m_e^{-2}$.
For example, when the elastic scattering of 
U(1) gauge field coupled to a light massive charged particle is considered, 
a lower bound on a charge-to-mass ratio analogous to the Weak Gravity Conjecture~\cite{Arkani-Hamed:2006emk} is obtained in this way~\cite{Cheung:2014ega}.
This analysis was extended
to more general setup such as the multiple U(1) gauge field case~\cite{Andriolo:2018lvp}, and various forms of Weak Gravity Conjecture are obtained or motivated~\cite{Andriolo:2018lvp,Hamada:2018dde, Chen:2019qvr,Arkani-Hamed:2021ajd,Noumi:2022ybv,Abe:2023anf,Bittar:2024xuc}.
One could also obtain bounds which are useful for phenomenological searches of dark sector physics~\cite{Noumi:2022zht,Aoki:2023khq,Kim:2024iud} (for earlier discussions along this line of considerations, see e.g., \cite{Andriolo:2018lvp,Alberte:2020jsk,Alberte:2021dnj}). 
The constraints however become weaker if negative $\sim m_e^{-2}$ contributions can be explained instead from the derivatives of the function $f(t)$ and $\alpha(t)$
entering \eqref{eq:sumrule_intro}.

Recent studies, where the graviton pole is treated in a more conservative way using for example smeared sum rules in place of forward limits, suggest that indeed these sign indefinite terms can be comparable to the large negative term found in \eqref{c2_eg1_intro}~\cite{Caron-Huot:2021rmr,Alberte:2021dnj,Noumi:2022wwf} (see also \cite{deRham:2022gfe,Hamada:2023cyt}).
This however only gives conservative bounds on $c_2(0)$ based on robust assumptions such as unitarity, analyticity, and crossing symmetry.\footnote{More precisely,
Regge behavior is used as an additional input in \cite{Noumi:2022wwf} where
it allows
to derive bounds without suffering from IR divergent issue in four spacetime dimensions.} 
A more precise evaluation of each term on the RHS of \eqref{eq:sumrule_intro} remains necessary in order to conclude whether the large negativity \eqref{c2_eg1_intro} of $c_2(0)$ is really allowed in a UV-complete quantum gravity setup such as string theory.

In this paper we study a string-inspired model meant to capture generic compactification in which the graviton is Reggeized and a light massive particle is present. We develop a method to precisely compute the $t$-dependence of $f(t)$ and $\alpha(t)$ at loop level, accounting for IR effects induced by exchange of a pair of light particles in the $t$-channel, in terms of EFT-calculable amplitudes together with Reggeized high-energy amplitudes.
We will focus on a $2\to2$ scalar elastic scattering as a proof of principle. Our main tool is the use of unitarity and discontinuity formulas to extract the $m^{-2}$-enhanced terms from a double discontinuity.

Applying our method to compute the $t$-dependence of our model stringy process, we will find large IR corrections to 
$\der_tf(t)$, but not to 
$\der_t\alpha(t)$ and $\der_t^2\alpha(t)$. We then find that the second term on the RHS in \eqref{eq:sumrule_intro} exactly reproduces the large negative term on the LHS (\eqref{c2_eg1_intro} for the case of photon scattering) calculated within EFT. 
This provides an example of an interesting interplay between low and high-energy processes in string amplitudes.
We do not consider this to be an example of
UV-IR mixing, however. In our view,
a better interpretation is
that physics in the regime $|t|\ll M_s^2\ll s$ is simply not UV physics.

This paper is organized as follows.
In section~\ref{sec:problem}, we review positivity bounds and the finite energy sum rules approach to dealing with the graviton pole, and introduce the stringy model to be analyzed.
In section~\ref{sec:single_regge}, we evaluate our stringy loop Regge amplitude at high energies and small-$t$ using unitarity and discontinuities,
and demonstrate the cancellation of negative term in the finite-energy sum rule. We discuss the robustness of 
this result.
In particular, in subsection \ref{sec:unitarity} we explain how two-particle threshold corrections to $f(t)$ and $\alpha(t)$ can be computed most readily using (an analytic continuation of) $t$-channel partial wave unitarity.
In section~\ref{sec:double_regge}, we confirm this perspective on $\alpha(t)$ 
working in the direct $s$-channel picture.
We conclude in section~\ref{sec:conclusion}. Several technical details are collected in appendices.

\section{Graviton exchange at loop level}\label{sec:problem}

\subsection{Motivation: Gravitational positivity}\label{sec:review}
Let us suppose that at low energies we have a QFT coupled to gravity in $D=4$ spacetime dimensions, which is UV-completed above the energy scale $\Ms$ by something resembling string theory.
For a while, let us suppose for simplicity
that the EFT consists of a light massive scalar field $\phi$ of mass $m$ in addition to the massless graviton. We also assume the presence of cubic self-interaction $\Lag\ni-gm\phi^3/6$, where $g$ denotes a dimensionless coupling constant. 
We will focus on the $\phi\phi\to \phi\phi$ scattering amplitude $\scat(s,t)$ where $(s,t)$ are usual Mandelstam variables. We ignore purely gravitational loops and work up to $\mathcal{O}(\Mpl^{-2})$ in the present analysis. 

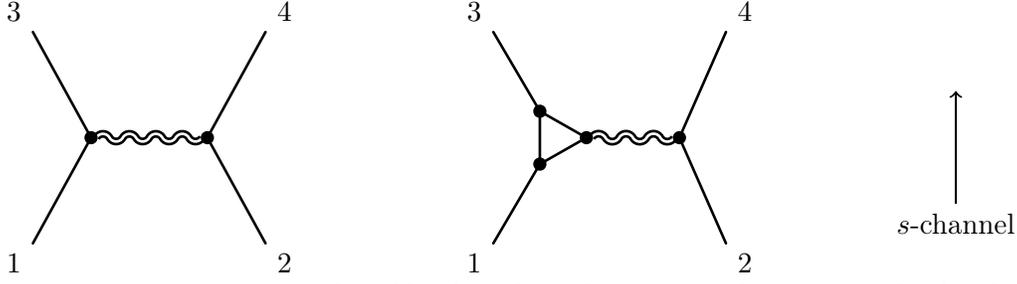
\begin{figure}[t]
\centering
\begin{fmffile}{A-B}
\begin{fmfgraph*}(110,80)
\fmfleft{i1,o3}
\fmfright{i2,o4}
\fmf{plain}{v1,o3}
\fmf{plain}{i1,v1}
\fmf{plain}{v2,o4}
\fmf{plain}{i2,v2}
\fmf{dbl_wiggly}{v1,v2}
\fmfdot{v1,v2}
\fmflabel{$1$}{i1}
\fmflabel{$2$}{i2}
\fmflabel{$3$}{o3}
\fmflabel{$4$}{o4}
\end{fmfgraph*}
\hspace{20mm}
\begin{fmfgraph*}(110,80)
\fmfleft{i1,o3}
\fmfright{i2,o4}
\fmf{plain}{v3,o3}
\fmf{plain}{i1,v1}
\fmf{plain}{v2,o4}
\fmf{plain}{i2,v2}
\fmf{plain}{v1,v3}
\fmf{plain}{v4,v1} 
\fmf{dbl_wiggly}{v2,v4}
\fmf{plain}{v3,v4}
\fmfdot{v1,v3,v2,v4}
\fmflabel{$1$}{i1}
\fmflabel{$2$}{i2}
\fmflabel{$3$}{o3}
\fmflabel{$4$}{o4}
\end{fmfgraph*}
\hspace{15mm}
\begin{tikzpicture}
    \draw[->,thick] (0,0) -- (0,1.5);
    \node[below] at (0,0) {$s$-channel};
\end{tikzpicture}
\end{fmffile}
\caption{$\phi\phi\to \phi\phi$ scattering mediated by the $t$-channel graviton exchange at tree-level and one-loop level, respectively. A solid line denotes a scalar field. Numbers attached to each external line label the external momentum: an external line with a number $j$ ($j=1,2,3,4$) has a momentum $p_j$. We use this notation throughout this paper.}
\label{fig:ph_dp_grav1}
\end{figure}

When the amplitude $\scat$ satisfies the Froissart-like found, $\lim_{|s|\to\infty}|\scat(s,t)/s^2|=0$ for $t<0$ as well as analyticity, one can derive sum rules for low-energy coefficients $\{c_n\}$ with $n\geq 2$ which are defined by \begin{align}
    \scat(s,t)|_\text{low-E}
    &=
    \text{(poles from light particles)}
    +
    \sum_{n=0}^\infty \frac{c_n(t)}{(2n)!}\left(\frac{s-u}{2}\right)^n
    \,.\label{eq:c2def}
\end{align}
For instance, the sum rule for $c_2(t)$ reads
\begin{align}
    c_2(t)
    =
        \frac{4}{\pi}\int^\infty_{\mth^2}\mathrm{d}s\,
        \frac{\disc_s\scat(s,t)}{(s-2m^2+t/2)^3}
        +
        \frac{2}{\Mpl^2t}
    \,,\label{eq:c2_sumrule0}
\end{align}
where the branch cuts at $s,-u-t\geq \mth^2$ account for the excitation of multi-particle states or heavy fields.
Calculable one-loop diagrams in the EFT,
such as the loop-corrected matter-matter-graviton vertex shown on the right in fig.~\ref{fig:ph_dp_grav1},
contribute negatively to $c_2(0)$ as~\cite{Alberte:2020jsk,Noumi:2021uuv} 
\begin{align}
    c_2(0)\supset
    -\frac{45-8\pi\sqrt{3}}{1296\pi^2}\frac{g^2}{\Mpl^2m^2}
    \approx -1.15\times 10^{-4}\frac{g^2}{\Mpl^2m^2}
    \,.\label{c2_eg1}
\end{align}
If \eqref{eq:c2_sumrule0} were to imply the condition $c_2(0)\geq 0$, then canceling the $m^{-2}$-enhanced negativity in \eqref{c2_eg1} would require the existence of additional low-energy physics. This gives nontrivial constraints on given models. These are sometimes closely related to the Weak Gravity Conjecture~\cite{Andriolo:2018lvp,Hamada:2018dde, Chen:2019qvr,Arkani-Hamed:2021ajd,Noumi:2022ybv,Abe:2023anf}, and sometimes are useful for phenomenological searches of dark sector physics~\cite{Noumi:2022zht,Aoki:2023khq,Kim:2024iud} (see also \cite{Andriolo:2018lvp,Alberte:2020jsk,Alberte:2021dnj}). 
Currently, no proof that $c_2(0)\geq 0$ is known. Given the phenomenological importance of the inequality, however, it is worthwhile investigating the sum rule for $c_2(0)$ in more detail, focusing on a scenario where the process on the right in fig.~\ref{fig:ph_dp_grav1} is also calculable at high energies.

In perturbative string theory, for instance, the assumed Regge boundedness is realized at tree-level through the Reggeization of graviton exchange at the tree-level,
$s^2\mapsto s^{2+\alpha't}$.
If it turns out that Reggeization occurs even at loop level in some range $\Ms^2\lesssim s \lesssim \Mstar^2$, then a more precise formula can be obtained by approximating $\scat(s,t)$ by the Regge amplitude $\scat_\text{R}(s,t)$ in the sum rule \eqref{eq:c2_sumrule0}, where the Regge amplitude is defined by
\begin{align}
    \scat_\text{R}(s,t)
    =
    \frac{-f(t)\left(1+e^{-i\pi\alpha(t)}\right)}{\sin(\pi\alpha(t))} \left(\frac{s}{\Ms^2}\right)^{\alpha(t)}
    \,,
    \quad
    \alpha(t)=2+\alpha't+\alpha''t^2/2+\cdots
    \,.
\end{align}
Notice that the imaginary part of $\scat_\text{R}(s,t)$ is regular as $t\to 0$: the pole from $\sin(\pi \alpha(t))$ is canceled by the imaginary part of the numerator (taking the leading trajectory to include the graviton so $\alpha(0)=2$).
Assuming Regge behavior to hold along a complex arc
along which $|s|\sim \Mstar^2$,
the portion of \eqref{eq:c2_sumrule0}
with $s>\Mstar^2$ can be replaced by the integral along that arc,
which nicely cancels the graviton pole given that the graviton is on the leading Regge trajectory (and corresponding constraint 
$\frac{2f(0)}{\pi M_s^4\alpha'(0)}=\frac{1}{\Mpl^2}$).
This leads to the following finite energy sum rule (FESR) for $c_2(0)$~\cite{Noumi:2022wwf}: 
\begin{align}
    c_2(0)
    \simeq
     \frac{4}{\pi}\int^{\Mstar^2}_{\mth^2}
    \mathrm{d}s
    \,
    \frac{\disc_s\scat(s,0)}{s^3}
    +
    \frac{1}{\Mpl^2}
    \left[
        -\frac{2\der_t\disc_s\scat_\text{R}}{\disc_s\scat_\text{R}}
        +
        \frac{\alpha''}{\alpha'}
    \right]_{t=0}
    \,.\label{eq:c2_sumrule}
\end{align}
It is important to stress that, even though the square bracket effectively replaces the integral of the forward limit of the amplitude over $s>\Mstar^2$, which formally admits a representation as a sum/integral of positive unknowns,
its sign \emph{cannot} be easily predicted from unitarity. This is because the sum and limit $t\to 0$ do not commute in the presence of massless gravitons and a divergent sum does not have a definite sign.  In fact the first term in the square bracket is negative due to unitarity.  This invalidates the simple positivity argument $c_2(0)\geq 0$ in the presence of a graviton pole.

Recent conservative treatments based on unitarity and analyticity suggest that the sign indefinite term can be comparable to the large negative term found in \eqref{c2_eg1}~\cite{Caron-Huot:2021rmr,Alberte:2021dnj,deRham:2022gfe,Noumi:2022wwf,Hamada:2023cyt}. It will be mandatory to evaluate them precisely, however, in order to investigate if the large negativity \eqref{c2_eg1} of $c_2(0)$ is really allowed. For this purpose, we must analyze the details of graviton Reggeization at loop level.

Note that the FESR assumes control of the amplitude for $|s|\lesssim \Mstar^2$ but does not require a Froissart-like bound at $s\to\infty$.  Conditions for the latter were discussed recently in \cite{Haring:2022cyf}.
Note also that since the low-energy contribution to the integral \eqref{eq:c2_sumrule} is EFT-calculable, one could further strengthen the sum rules by moving part of the integral to the left-hand side~\cite{Bellazzini:2016xrt,deRham:2017imi}.
We will not pursue this here because the imaginary part of the considered diagram in fig.~\ref{fig:ph_dp_grav1} vanishes within the EFT.

Throughout this paper we assume that $\frac{g^2}{\Mpl^2m^2}\gg \frac{1}{\Mpl^4}$, so that purely gravitational loops can be neglected. In other words, we only consider situations in which the Weak Gravity Conjecture is parametrically satisfied. The main motivation for this is phenomenology.  We will see that even in situations where graviton loops can be neglected, the square bracket of \eqref{eq:c2_sumrule} can in fact be large and negative. 

\subsection{String amplitudes and channel duality}\label{sec:dual}

\begin{figure}[t]
\centering
\begin{fmffile}{treeRegge_closed} 
\begin{align*}
\raisebox{-0.45\height}{
\begin{fmfgraph*}(100,70)
\fmfleft{i1,m1,o3}
\fmfright{i2,m2,o4}
\fmf{plain,label.side=left}{v2,o3}
\fmf{plain,label.side=left}{i1,v1}
\fmf{plain,label.side=right}{v2,o4}
\fmf{plain,label.side=right}{i2,v1}
\fmf{dbl_plain}{v1,v2}
\fmfdot{v1,v2}
\fmf{dashes,fore=red}{m1,m2}
\fmflabel{$1$}{i1}
\fmflabel{$2$}{i2}
\fmflabel{$3$}{o3}
\fmflabel{$4$}{o4}
\end{fmfgraph*}
}
\qquad
\equiv
\qquad
\raisebox{-0.45\height}{
\begin{fmfgraph*}(110,70)
\fmfleft{i1,m1,o3}
\fmfright{i2,m2,o4}
\fmf{plain,label.side=left}{v1,o3}
\fmf{plain,label.side=left}{i1,v1}
\fmf{plain,label.side=right}{v2,o4}
\fmf{plain,label.side=right}{i2,v2}
\fmf{dbl_wiggly,tension=-0.1,label=$g_*$}{v1,v2}
\fmfdot{v1,v2}
\fmf{dashes,fore=red}{m1,m2}
\fmf{dashes,fore=red}{v1,v2}
\fmflabel{$1$}{i1}
\fmflabel{$2$}{i2}
\fmflabel{$3$}{o3}
\fmflabel{$4$}{o4}
\end{fmfgraph*}
}
\quad
\raisebox{-0.45\height}{
\begin{tikzpicture}
    \draw[->,thick] (0,0) -- (0,1.5);
    \node[below] at (0,0) {$s$-channel};
\end{tikzpicture}
}
\end{align*}
\end{fmffile}
\caption{Channel duality for $\disc_s\scat_\text{type II}(s,t)$: summing over $s$-channel exchanges of heavy higher-spin particles is equivalent to the $t$-channel exchange of a Reggeized graviton, as shown on the left and right, respectively.  A red dashed line indicates that the $S$-channel discontinuity is evaluated.
Throughout this paper, a double solid line represents a collection of heavy stringy states while a double wiggly line (labelled $g_*$ here)
denotes a Reggeized graviton.}
\label{fig:treeRegge_closed1}
\end{figure}
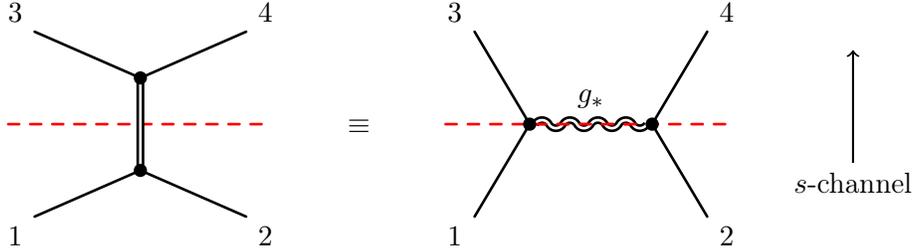

How is the energy growth of graviton exchange amplitudes tamed in quantum gravity?  
Tree-level string theory offers simple examples. Consider the four dilaton amplitude in type-II closed string theory.
The Reggeization of the $t$-channel graviton exchange is realized through $s,u$-channel exchanges of heavy higher-spin towers which are excitations of heavy closed string. This amplitude is represented by the diagrams shown in fig.~\ref{fig:treeRegge_closed1}. Explicitly, the four dilaton amplitude at tree-level is given by the Virasoro-Shapiro formula
\begin{subequations}
    \label{eq:type2_amp}
\begin{align}
&   \scat_\text{type II}(S,T)
    =
    \left.
    -P(S,T)\frac{\Gamma(-S/4)\Gamma(-T/4)\Gamma(-U/4)}{\Gamma(1+S/4)\Gamma(1+T/4)\Gamma(1+U/4)}
    \right|_{U=-S-T}
    \,,\\
&   P(S,T)
   \equiv 
   \frac{1}{64\Mpl^2}
   (S^2U^2+T^2U^2+S^2T^2)_{U=-S-T}
    \,.
\end{align}
\end{subequations}
Here, $(S,T,U)$ are Mandelstam variables defined in $D=10$ dimensions and satisfying $S+T+U=0$. We set $\Ms=2$, $\alpha'=1/2$ and $8\pi G=\Mpl^{-2}$. The discontinuity in the $S$-channel physical region is 
\begin{subequations}
\label{eq:type2_disc}
\begin{align}
&    \left.
    \disc_S\scat_\text{type II}(S,T)
    \right|_{S\geq 0,T\sim0}
    = \sum_{n=0}^\infty I_n(S,T)
    \,,
    \\
&   I_n(S,T)
    \equiv
    \pi P(S,T)
    \left(
        \frac{\Gamma(n+T/4)}{\Gamma(1+T/4)\Gamma(1+n)}
    \right)^2
    \delta(S/4-n)  \label{eq:type2_disc_b}
    \,.
\end{align}
\end{subequations}
The lightest higher-spin state appearing in this amplitude has spin $4$ and mass $\Ms$. 
This amplitude exhibits the Regge behavior in the domain $S\gg\Ms^2,|T|$:
\begin{align}
    \label{eq:type2_regge}
 \disc_s\scat_\text{type II}(S,T) 
  &\simeq f_\text{VS}(T) \left(\frac{S}{\Ms^2}\right)^{\alpha(T)}
\end{align}
with the coefficient and Regge trajectory
\begin{align} \label{eq:fVS}
  f_\text{VS}(T) = \frac{1}{\Mpl^2}
  \frac{\pi}{\left[\Gamma(1+\alpha'T/2)\right]^2},\quad
 \alpha(T)=2+\alpha'T \quad\mbox{with } \alpha'=1/2.
\end{align}
This is a manifestation of channel duality: the Regge limit of $\disc_s\scat_\text{type II}(S,T)$ can be described equivalently in terms of exchange of a Reggeized graviton in $T$-channel, \emph{or} in terms of many
heavy $S$-channel states. 

A technical comment is in order. The approximation \eqref{eq:type2_regge} fails pointwise because it misses the $\delta$-function in \eqref{eq:type2_disc_b}. Nonetheless, it holds in an average sense
\cite{Collins:1977jy}. That is,
it holds whenever integrated against a sufficiently smooth function of $S$. A convenient definition of the averaged discontinuity is as the difference between amplitudes rotated a finite amount into the complex plane, ie.
\begin{align}
 \disc_s\scat(S,T) \longrightarrow \frac{1}{2i}
\Big[ \scat(S(1+i\epsilon),T)-\scat(S(1-i\epsilon),T)\Big]
\end{align}
where one keeps the angle $\epsilon$ finite.
This smears out the infinite set of $\delta$-functions, for example
\begin{align}
\sum_n \delta(S/4-n) = \disc_s\,\cot(-\pi S/4)
\longrightarrow \frac{1}{2i}\Big[\cot(-\pi S/4(1+i\epsilon))
-\cot(-\pi S/4(1-i\epsilon))\Big] =1.
\end{align}
All high-energy approximations in this paper should be understood to hold in this averaged sense. This is sufficient for understanding dispersive integrals like
\eqref{eq:c2_sumrule0}.

At loop level, we would like to understand 
how exchange of string states shown fig.~\ref{fig:loopRegge_closed1} may be responsible for Reggeizing the low-energy loop process represented by fig.~\ref{fig:ph_dp_grav1}.
We analyze this type of diagram quantitatively in section~\ref{sec:single_regge} in light of the $c_2$-sum rule \eqref{eq:c2_sumrule}.%
\footnote{In fact, there are various type of Regge amplitude for the tree-level graviton exchange other than \eqref{eq:type2_amp}. We briefly comment on how our analysis may be extended to such more general cases in section~\ref{sec:unitarity}.}

\subsection{Massive string amplitude}\label{sec:setup}

The essential features of the string loop diagrams we would like to study are the presence of a massless graviton and of light massive particle,
which together lead to the large negative term \eqref{c2_eg1} enhanced by $1/m$.
We will consider light massive scalars arising from a Kaluza-Klein compactification.
We will assume that the tree amplitudes involving these particles are given by the Virasoro-Shapiro formula, but now expressed in terms of momenta $P_i=(p_i^\mu,p_i^\perp)$ which include both four-dimensional and higher-dimensional component. The square of extra-dimensional momenta are interpreted simply as four-dimensional masses:
\begin{align}
(P_i)^2 =0 \quad\Leftrightarrow\quad m_i^2 = -p_i^2 = (p_i^\perp)^2\,.
\end{align}
The higher-dimensional Mandelstam invariants are given as
\begin{align}
 S=-(p_1+p_2)^2 - (p_1^\perp+p_2^\perp)^2 \equiv s-m_S^2\,,
\end{align}
and similarly for the $T$ and $U$ channels.
This allows us to use the Virasoro-Shapiro amplitude to describe for a generic scattering $m_1,m_2\to m_3,m_4$: 
\begin{align} \label{VS with masses}
 \scat_{\{m_i\}}(s,t) =
\scat_\text{type II}(S=s-m_S^2,T=t-m_T^2)\Big|_{
 U=u-m_U^2}
 \,.
\end{align}

There can be various angles between
the extra-dimensional momenta but the internal and external masses must always be related as:
\begin{align} m_S^2+m_T^2+m_U^2=\sum_{i=1}^4 m_i^2,
\label{mass constraint}
\end{align}
which ensures $S+T+U=0$ and is simply interpreted as momentum conservation in the extra dimensions.
Thus, the amplitudes we consider are functions of six independent mass squared.
Requiring all $p_i^\perp$ to be real would further restrict the allowed masses, but we will ignore this constraint.

At loop level, in an actual toroidal compactification we would find a discrete sum over Kaluza-Klein momenta.
These would correspond to a tower of particles with mass comparable to that of the light scalar.  Instead, we will assume that the compactification under study leads to a few scalars which are parametrically lighter than all other modes.
In effect we will select the contribution of desired states through unitarity cuts. Our main assumption will be that we can describe the corresponding amplitudes using
the Regge ansatz \eqref{eq:type2_regge} together with
the substitution in \eqref{VS with masses}.

\subsection{Setup: string loops and double discontinuity}\label{stringloop}

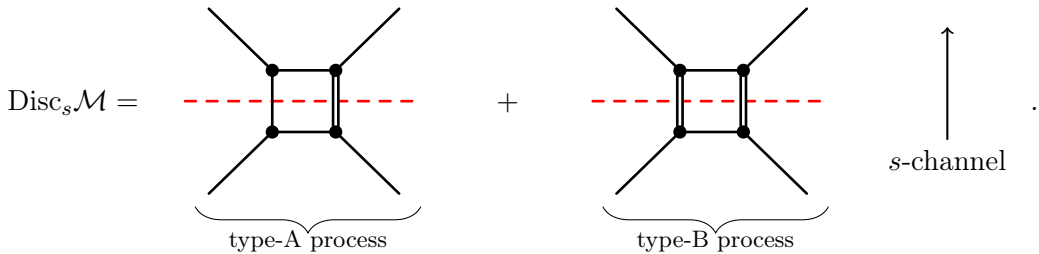
\begin{figure}[tbp]
\centering
\begin{fmffile}{loopRegge_closed} 
\begin{align*}
\disc_s\scat
=
\quad
\raisebox{-0.45\height}{
\begin{fmfgraph*}(90,70)
\fmfleft{i1,m1,o3}
\fmfright{i2,m2,o4}
\fmf{plain,label.side=left}{v3,o3}
\fmf{plain,label.side=left}{i1,v1}
\fmf{plain,label.side=right}{v4,o4}
\fmf{plain,label.side=right}{i2,v2}
\fmf{plain}{v1,v3}
\fmf{plain}{v2,v1} 
\fmf{dbl_plain}{v2,v4}
\fmf{plain}{v3,v4}
\fmfdot{v1,v3,v2,v4}
\fmf{dashes,fore=red}{m1,m2}
\raisebox{-0.15\height}{
\begin{tikzpicture}
    \node (n) at (0,0) {};
    \draw [decorate,decoration={brace,amplitude=10pt,raise=4pt}]
    (-7,-5) -- (-10,-5) node [black,midway,yshift=-0.6cm] {\footnotesize type-A process};
\end{tikzpicture}
}
\end{fmfgraph*}
}
\qquad
+
\qquad
\raisebox{-0.45\height}{
\begin{fmfgraph*}(90,70)
\fmfleft{i1,m1,o3}
\fmfright{i2,m2,o4}
\fmf{plain,label.side=left}{v3,o3}
\fmf{plain,label.side=left}{i1,v1}
\fmf{plain,label.side=right}{v4,o4}
\fmf{plain,label.side=right}{i2,v2}
\fmf{dbl_plain}{v1,v3}
\fmf{plain}{v2,v1} 
\fmf{dbl_plain}{v2,v4}
\fmf{plain}{v3,v4}
\fmfdot{v1,v3,v2,v4}
\fmf{dashes,fore=red}{m1,m2}
\raisebox{-0.15\height}{
\begin{tikzpicture}
    \node (n) at (0,0) {};
    \draw [decorate,decoration={brace,amplitude=10pt,raise=4pt}]
    (-7,-5) -- (-10,-5) node [black,midway,yshift=-0.6cm] {\footnotesize type-B process};
\end{tikzpicture}
}
\end{fmfgraph*}
}
\quad
\raisebox{-0.45\height}{
\begin{tikzpicture}
    \draw[->,thick] (0,0) -- (0,1.5);
    \node[below] at (0,0) {$s$-channel};
\end{tikzpicture}
}
\,.
\end{align*}
\end{fmffile}
\vspace{2mm}
\caption{
Two types of contributions to the one-loop unitarity relation at high energies: either one of the states on the cut is light (type-A), or both states are heavy (type-B).
Focusing on the double-discontinuity $\rho_{st}$ at small $t$ will force the horizontal lines to be also on-shell and light (with complexified momenta since $t>0$),
thus reducing the calculation to a sum of field-theoretic box diagrams.
}
\label{fig:loopRegge_closed1}
\end{figure}

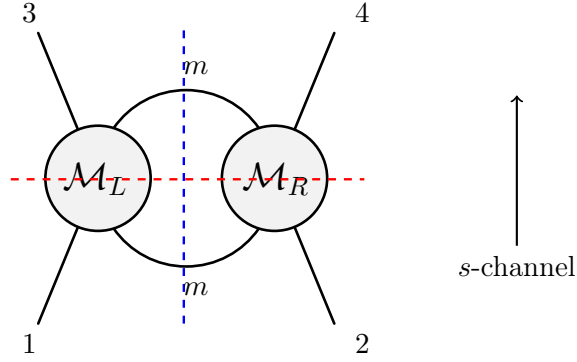
\begin{figure}[tbp]
\centering
\begin{fmffile}{general_box} 
\begin{fmfgraph*}(140,110)
\fmfleft{i1,m1,i2}
\fmfright{o1,m2,o2}
\fmf{plain,label.side=left}{i2,v1}
\fmf{plain,label.side=left}{i1,v1}
\fmf{plain,label.side=left}{v2,o2}
\fmf{plain,label.side=left}{v2,o1}
\fmf{plain,label=$m$,label.side=above,left=1}{v1,v2,v1}
\fmf{phantom,tension=-1.5}{v1,v2}
\fmf{phantom,tension=1.5}{v1,m1}
\fmf{phantom,tension=1.5}{v2,m2}
\fmf{phantom,fore=red}{m1,m2}
\fmf{phantom,fore=red}{v1,v2}
\fmfv{decor.shape=circle,decor.filled=5,decor.size=14mm}{v1,v2}
\fmflabel{$1$}{i1}
\fmflabel{$2$}{o1}
\fmflabel{$3$}{i2}
\fmflabel{$4$}{o2}
\begin{tikzpicture}[overlay]
    \node at (1.25,1.92) {\Large $\mathcal{M}_L$}; 
    \node at (3.65,1.92) {\Large $\mathcal{M}_R$};
\end{tikzpicture}
\begin{tikzpicture}[overlay]
     \draw[blue, dashed, line width=0.9pt] (2.3,0) -- (2.3,4);
     \draw[red, dashed, line width=0.9pt] (0,1.92) -- (4.7,1.92);
\end{tikzpicture}
\end{fmfgraph*}
\qquad
\raisebox{0.2\height}{
\begin{tikzpicture}
    \draw[->,thick] (0,0.5) -- (0,2.5);
    \node[below] at (0,0.5) {$s$-channel};
\end{tikzpicture}
}
\end{fmffile}
\caption{The $t$-channel discontinuity we will consider in sections \ref{sec:single_regge} and \ref{sec:double_regge} is represented by the blue dashed line. The two exchanged particles are light (mass $m$) while
the blobs will be either tree-level exchanges at low energies, or string amplitudes at high energies.
}
\label{fig:general_box}
\end{figure}

Despite continuing advances in calculations of string loop amplitudes (see for example \cite{Banerjee:2024ibt} and references therein), explicit calculations for a generic compactification remain challenging.
The following considerations will greatly simplify our task:
\begin{enumerate}
\item We are only interested in the imaginary part
$\disc_s\scat(s,t)$, which can be computed by unitarity as a product of tree amplitudes.
\item We will focus on $1/m$-enhanced terms in the forward limit or its derivative.
\end{enumerate}
The imaginary part is simpler to calculate than the full amplitude because unitarity expresses it as a product of string trees; this has been used in the string context recently in \cite{Eberhardt:2022zay}.
However, one still has to sum over the polarizations of heavy intermediate states of arbitrary spin, which products nontrivial
polynomials in Mandelstam invariants.
However, we will argue in the next section that $1/m$-enhanced terms can only come from singularities near $t=0$, which are captured by the double discontinuity:
\begin{align}
    \rho_{st}\equiv {\rm Disc}_t{\rm Disc}_s \scat(s,t)\,.
\end{align}
Only the singularity closest to $t=0$ will matter to us, which will single out light particles exchanged in the \emph{$t$-channel}.
This reduces the calculation to a sum of field-theoretic quadruple cuts (box diagrams), where the numerator at each mass level is readily determined using $t$-channel factorization.
In order to relate with the field theory diagram in fig.~\ref{fig:ph_dp_grav1}, we will focus on contribution with two scalars of mass $m$ in the $t$-channel, see fig.~\ref{fig:general_box}.

The following formula will be used to calculate the double discontinuity of a generic box diagram:
\begin{align}
   \rho_{st}(s,t)
    =\frac{1}{8\pi^2\sqrt{t(t-4m^2)}}
    \int^{s}_0\mathrm{d}s_L\int^{s_-(s,t;s_L)}_0\mathrm{d}s_R
\frac{\disc_{s_L}\scat_L(s_L,t)\disc_{s_R}\scat_R^*(s_R,t)}{(s_R-s_-)^{1/2}(s_R-s_+)^{1/2}}
    \,,
    \label{eq:formula_doubledisc}
\end{align}
where $\scat_L$ and $\scat_R$ denote sub diagrams on the left and right, respectively, and $s_\pm(s,t;s_L)$ is defined by
\begin{align}
    s_\pm(s,t;s_L)
&   \equiv
    \frac{
    s \left(2 s_L+t-4m^2\right)
    \pm
    2 \sqrt{s s_L \left(s+t-4 m^2\right) \left(s_L+t-4 m^2\right)}
    }
    {t-4 m^2}
    +s_L
    \,.\label{eq:range}
\end{align} 
See appendix~\ref{sec:doubledisc_derivation} for a brief summary of derivation of \eqref{eq:formula_doubledisc}. We will use \eqref{eq:formula_doubledisc} in succeeding sections. From eq.~\eqref{eq:formula_doubledisc}, we obtain the support of $\rho_{st}(s,t)$ for a given diagram. 
In particular, at large $s$, $\rho_{st}$ vanishes unless $t>4m^2$.

\begin{figure}[tbp]
 \centering
\includegraphics[width=0.9\textwidth, trim=110 140 40 160,clip]{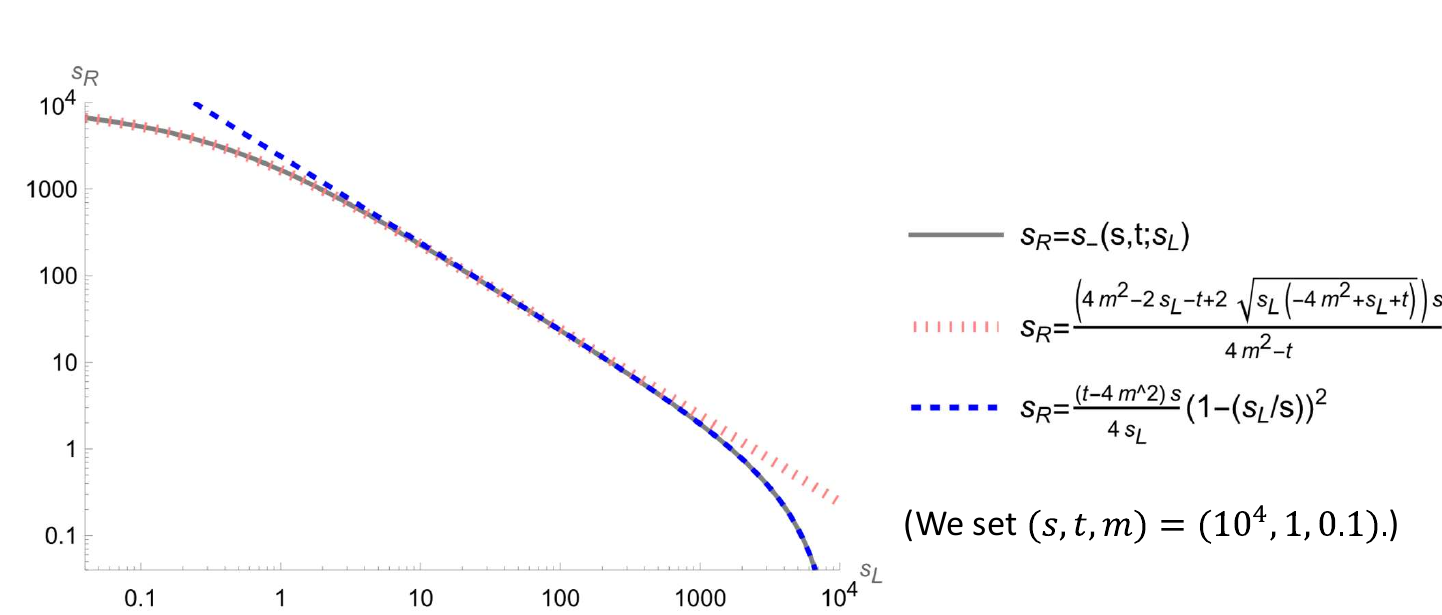}
 \caption{The integration domain in \eqref{eq:formula_doubledisc} 
is the part of the first quadrant which lies below the solid gray curve.  In this plot we set $(s,t,m)=(10^4,1,0.1)$ for illustration.  The approximations \eqref{eq:range_est1} and \eqref{eq:range_est2} are also shown as the blue dotted line
and pink dashed line, respectively.
}
 \label{fig:int_dom_1} 
\end{figure}

We display the line $s_R\leq s_-(s,t;s_L)$ in the $(s_L,s_R)$-plane in fig.~\ref{fig:int_dom_1}, which bounds the domain of integration of \eqref{eq:formula_doubledisc}.
In the regime $s,s_L\gg t,m^2$, $s_\pm(s,t;s_L)$ is well approximated as 
\begin{subequations}
\label{eq:range_est}
\begin{align}
    &s_-(s,t;s_L)
    =
    \frac{(t-4m^2) s}{4s_L}
    \left(1-(s_L/s)\right)^2
    +\mathcal{O}\left(\frac{(t,m^2)}{(s,s_L)}\right)
    \,,\label{eq:range_est1}
    \\
    &s_+(s,t;s_L)
    =
    \frac{4s_Ls}{t-4m^2}
     +\mathcal{O}\left(\frac{(t,m^2)}{(s,s_L)}\right)
    \,.
\end{align}
\end{subequations}
Permuting $s_L$ and $s_R$ we get equivalently in the region $(t, m^2 , s_L) \ll s$:
\begin{align}
    s_\pm(s,t;s_L)
    =
    \frac{
    s
    \left(
    t+2 s_L -4 m^2
    \pm
    2 \sqrt{s_L \left(s_L+t-4 m^2\right)}
    \right)
    }
    {t-4m^2}
    +
    \mathcal{O}(s^0)
    \,.
    \label{eq:range_est2}
\end{align}
When the two cut particles are heavy (string states), \eqref{eq:range_est1} suggests that
$\rho_{st}$ is only nonzero for $s>\frac{4s_L s_R}{t-4m^2}\sim \frac{\Ms^4}{t-4m^2}$.
This means that to get a singularity at small $t$, we need either one $s$-channel state $s_L$ or $s_R$ to be light, or the total center-of-mass energy to be sufficiently large.

We are interested in energies $M_s^2 \lesssim s \lesssim \Mstar^2$ which are above the EFT cutoff but within the Regge regime.
For each discontinuity $\disc_s\scat_{L/R}$, we will have to distinguish the cases in which the cut particles of energy $s_L$ and $s_R$ are light or heavy. We refer to these as the $A$ and $B$ contributions, see fig.~\ref{fig:loopRegge_closed1}. We treat them separately in the next two sections.

When the cut particle $s_i$ is light, we use in \eqref{eq:formula_doubledisc} the (discontinuity) of the EFT amplitude, and when it is heavy we use the Virasoro-Shapiro formula \eqref{eq:type2_amp} with appropriate mass substitutions
\eqref{VS with masses}, which will be detailed as we go.
While we will use the full Virasoro-Shapiro amplitudes in our calculations, we will find that the contribution of towers are essentially determined by the Regge approximation \eqref{eq:type2_regge}. Thus, our essential assumption is that the true tree-level amplitudes $\scat_L$ and $\scat_R$ interpolate between the EFT form at low energies and Regge behavior above the string scale.

To summarize, we consider a string-inspired setup in which we include a subset of the one-loop unitarity contributions in a toroidal compactification, defined by selecting specific states on the cut.
We also assume that at low energies the amplitude on the left involves a dimensionless scalar cubic coupling $g$ smaller than unity but sufficiently large that
we can ignore $\mathcal{O}(\Mpl^{-4})$ terms compared with 
the effects $\sim \frac{g^2}{\Mpl^2m^2}$ that we focus on
(we comment on the case of smaller $g$ in conclusion).
While it would be interesting to consider genuine string theory compactifications, we believe that this setup captures the main physics of compactifications in which parametrically light particles appear.

\section{Loop amplitudes involving a single Regge tower}
\label{sec:single_regge}

In this section we analyze contributions to the double discontinuity $\rho_{st}$ arising
from a light particle in the left amplitude and a Reggeized tower on the right, as shown in fig.~\ref{fig:box_closed_1}.
In subsection \ref{sec:Disc_sumrule} we use a dispersive argument to ``undo'' the $t$-channel discontinuity and infer the contribution to ${\rm Disc}_s$, which we confront with the finite-energy sum rule (FESR) introduced above.  
In subsection~\ref{sec:unitarity} we explain how these results can be understood most directly using unitarity of (analytically continued) $t$-channel partial waves, which we argue makes our results robust when considering other Regge amplitudes.

\subsection{Regge behavior of double discontinuity}\label{closed_model1}

\begin{figure}[tbp]
 \centering
  \includegraphics[width=1\textwidth, trim=10 100 10 220,clip]{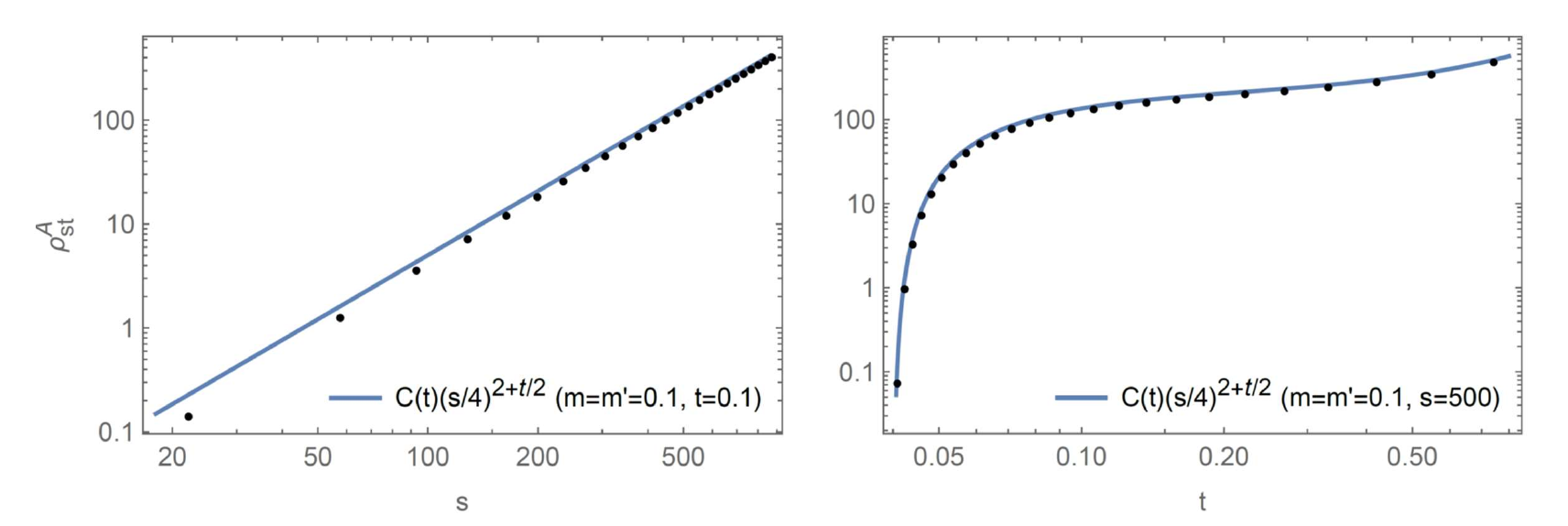}
  \caption{Values of $\rho^A_{st}(s,t)$ are plotted for fixed $t=0.1$ and $s=500$ on the left and right panel, respectively. We set $\Mpl=g=1$ and choose $m=m'=0.1$ in these plots. Black dots show the numerical evaluations of the exact sum \eqref{eq:jsum}: on the left(right) panel, we choose the values of $s$ for which we have $s_*^-(s,0.1)=4n+2$ $(s_*^-(500,t)=4n+2)$ with $n=2,6,10,\cdots$. A blue solid line on each panel represents the analytic estimate \eqref{eq:type2_regge_estimate1}. In both cases, the exact results agree well with the analytic estimate at large $s\gg \Ms^2$.
}
\label{fig:fixedst_reggefit} 
\end{figure}

The contribution to $\rho_{st}$ under consideration is equal, using $t$-channel factorization, to a product of an EFT-computable scalar exchange amplitude $\scat_\text{scalar} \propto (s_L-{m'}^2)^{-1}$ and a type II high-energy amplitude $\scat_\text{type II}(S,T)$.  More precisely, we require the discontinuity of these ingredients:
\begin{align}
\disc_s\scat_L(s,t)&=(gm)^2\pi\delta(s-{m'}^2),\\
\disc_s\scat_R(s,t)&=\disc_s\scat_\text{type II}(S=s-2{m}^2,t).
\end{align}
Here for the stringy amplitude we have chosen mass shifts in \eqref{VS with masses} such that $m_T^2=0$ (we want a massless graviton) while maintaining $S{\leftrightarrow}U$ symmetry and $S+T+U=4m^2$. However the precise choice will not 
affect the calculations below as long as $m^2\ll \Ms^2$.
Using \eqref{eq:type2_disc} and inserting into \eqref{eq:formula_doubledisc}:
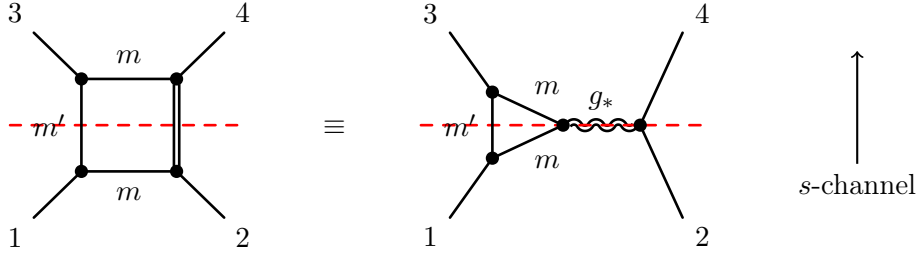
\begin{figure}[tbp]
\centering
\begin{fmffile}{scalar_fourpoint} 
\begin{align*}
\raisebox{-0.45\height}{
\begin{fmfgraph*}(90,70)
\fmfleft{i1,m1,i2}
\fmfright{o1,m2,o2}
\fmf{plain,label.side=left}{i2,v2}
\fmf{plain,label.side=left}{i1,v1}
\fmf{plain,label.side=left}{v4,o2}
\fmf{plain,label.side=left}{v3,o1}
\fmf{plain,label=$m'$,label.side=left,tension=0.5}{v1,v2}
\fmf{plain,label=$m$,label.side=left,tension=0.5}{v2,v4} 
\fmf{dbl_plain,tension=0.5}{v3,v4}
\fmf{plain,label=$m$,label.side=left,tension=0.5}{v3,v1}
\fmfdot{v1,v2,v3,v4}
\fmflabel{$1$}{i1}
\fmflabel{$3$}{i2}
\fmflabel{$2$}{o1}
\fmflabel{$4$}{o2}
\fmf{dashes,fore=red}{m1,m2}
\end{fmfgraph*}
}
\qquad
\equiv
\qquad
\raisebox{-0.45\height}{
\begin{fmfgraph*}(110,70)
\fmfleft{i1,m1,o3}
\fmfright{i2,m2,o4}
\fmf{plain,label.side=left}{v3,o3}
\fmf{plain,label.side=left}{i1,v1}
\fmf{plain,label.side=right}{v2,o4}
\fmf{plain,label.side=right}{i2,v2}
\fmf{plain,label=$m'$,label.side=left,tension=0.6}{v1,v3}
\fmf{plain,label=$m$,tension=0.6}{v4,v1} 
\fmf{dbl_wiggly,label=$g_*$,tension=0.1}{v2,v4}
\fmf{plain,label=$m$,label.side=left,tension=0.6}{v3,v4}
\fmfdot{v1,v3,v2,v4}
\fmf{dashes,fore=red}{m1,m2}
\fmf{dashes,fore=red}{v2,v4}
\fmflabel{$1$}{i1}
\fmflabel{$2$}{i2}
\fmflabel{$3$}{o3}
\fmflabel{$4$}{o4}
\end{fmfgraph*}
}
\qquad
\raisebox{-0.45\height}{
\begin{tikzpicture}
    \draw[->,thick] (0,0) -- (0,1.5);
    \node[below] at (0,0) {$s$-channel};
\end{tikzpicture}
}
\end{align*}
\end{fmffile}
\caption{Channel duality for $\disc_s\scat_A$ for type-A contributions, where one of the $s$-channel particles has a small mass $m'$. All external lines have mass $m$ and the solid double line represents an infinite sum over heavy string states with mass
$n\Ms^2+2m^2$.  Since these heavy states are dual to the Reggeized graviton, we expect the sum to be well approximated by the picture on the right.
}
\label{fig:box_closed_1}
\end{figure}

\begin{align}
&    \rho_{st}^A(s,t)
    \equiv \disc_t\disc_s\scat_A(s,t)
    = \sum_{n=1}^\infty \rho_{n}(s,t)\,,\label{eq:jsum}
    \\
&    \rho_{n}(s,t)
    =\frac{(gm)^2}{8\pi \sqrt{t(t-4m^2)}}
    \int^\infty_{0}\mathrm{d}s_R\, I_n(s_R{-}2m^2,t)\,\frac{\Theta(s_*^-(s,t)-s_R)}{\sqrt{(s_R-s_*^+(s,t))(s_R-s_*^-(s,t))}}
    \,,\label{eq:0j_type2model1}
\end{align}
where $s_*^\pm(s,t)\equiv s_\pm(s,t;{m'}^2)$. 
We can simply substitute the exact expression of $I_n$ in \eqref{eq:type2_disc} and numerically perform the summation over $n$ to evaluate $\rho_{st}^A(s,t)$ for given $s$.
The results are plotted in fig.~\ref{fig:fixedst_reggefit}. 

For a while, we focus on the configuration $s\gg \Ms^2\gg t$ since we are particularly interested in the threshold singularity at $t\sim 4m^2 \ll \Ms^2$ for the gravitational Regge amplitude. 
In this domain, we can approximate $\rho_{st}^A(s,t)$ using the Regge behavior recorded in \eqref{eq:fVS}: 
\begin{align}
    \rho_{st}^A(s,t)
    &\approx
    \frac{f_\text{VS}(t)(gm)^2}{8\pi\sqrt{t(t{-}4m^2)}}
    \int^\infty_0\mathrm{d}s_R\,
    \frac{(s_R/4)^{\alpha_\tree(t)}\Theta(s_*^-{-}s_R)}{\sqrt{(s_R{-}s^+_*)(s_R{-}s^-_*)}}
    =
    C(t)
    \left(\frac{s}{4}\right)^{\alpha_\tree(t)}
    \left[1{+}\mathcal{O}\left(s^{-1}\right)\right]
    \,,\label{eq:type2_regge_estimate1}
\end{align}
with
\begin{align}
    C(t)
    &=
    \frac{(gm)^2f_\VS(t)}{8\pi \sqrt{t(t-4m^2)}}
    Q_{\alpha_\tree(t)}\left(1+\frac{2{m'}^2}{t-4m^2}\right)
    \Theta(t-4m^2)
    \,.\label{Cdef_general}
\end{align}
We describe the computational details to obtain eqs.~\eqref{eq:type2_regge_estimate1} and \eqref{Cdef_general} in appendix~\ref{sec:compute_C}. Instead, here we make a few observations on \eqref{Cdef_general}. First, the fact that the $s_R$ integral yields a Legendre-Q function $Q_{\alpha_\tree}$ is intriguing: this is related to the $t$-channel partial wave amplitude of $\scat_L$,
\begin{align}
    \frac{1}{32\pi}\int^1_{-1}\mathrm{d}z_s\,
    \scat_L(s,t)P_J(z_s)
    &= 
    \frac{(gm)^2}{16\pi(t-4m^2)}\int^1_{-1}\mathrm{d}z_s\,
    \frac{P_J(z_s)}{z_{{m'}^2}-z_s}
    =
    \frac{(gm)^2Q_J\left(z_{{m'}^2}\right)}{8\pi(t-4m^2)}
    \,,
\end{align}
where $z_s\equiv 1+2s/(t-4m^2)$.
In effect, at high energies the $s_R$ integral appears to reduce to the Froissart-Gribov formula projecting the EFT amplitude onto its component of (continuous) spin $j=\alpha_\tree(t)$.
Hence, the $t$-channel factorization is inherent in \eqref{Cdef_general}. We will see how it can be directly obtained using $t$-channel unitarity in subsection~\ref{sec:unitarity}.

Substituting $f_\text{VS}(t)$ and $\alpha_\tree(t)$ for the type-II amplitude, we obtain
\begin{align}
    C(t)
    &=
    \frac{(gm)^2}{\Mpl^2}
    \frac{1}{\left[\Gamma(1+t/4)\right]^2}
    \frac{1}{2 \sqrt{t(t-4m^2)}}
    Q_{2+t/2}\left(1+\frac{2{m'}^2}{t-4m^2}\right)
    \Theta(t-4m^2)
    \,.\label{eq:c_def1}
\end{align}
We can check that \eqref{eq:type2_regge_estimate1} with \eqref{eq:c_def1} is a good approximation for the exact sum over residues at large $s$ for various values of $t$: see fig.~\ref{fig:fixedst_reggefit}.

\begin{figure}[tbp]
 \centering
  \includegraphics[width=.5\textwidth, trim=0 0 0 0,clip]{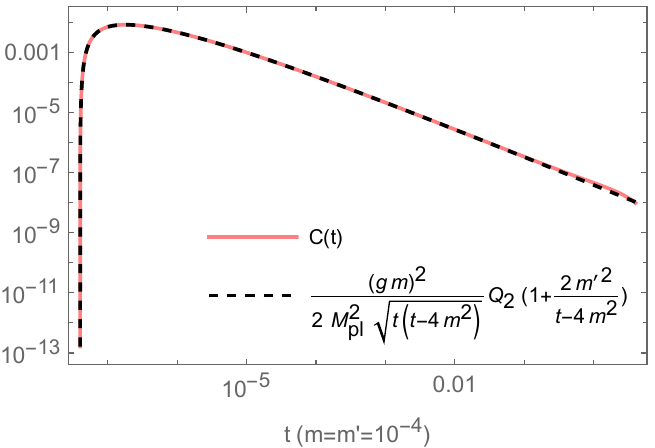}
 \caption{A plot of $C(t)$ for given $m=m'=10^{-4}$ is shown in the solid line. We set $\Mpl=g=1$ in this plot. An approximate expression \eqref{eq:C_app1} is shown in the dashed line, which agrees well with the solid line at $t\leq 1$.}
 \label{fig:ct_single_shape0} 
\end{figure}

At small $t\ll \Ms^2$, the function $C(t)$ can be further approximated as 
\begin{align}
     &C(t)
     \simeq
     \frac{(gm)^2}{\Mpl^2}
    \frac{1}{2\sqrt{t(t-4m^2)}}
    Q_{2}\left(1+\frac{2{m'}^2}{t-4m^2}\right)
    \Theta(t-4m^2)
    \qquad
    (t\ll\Ms^2)
    \,.
    \label{eq:C_app1}
\end{align}
Plots of $C(t)$ based on
the approximation \eqref{eq:C_app1} are shown in fig.~\ref{fig:ct_single_shape0}. The figure confirms that the approximation \eqref{eq:C_app1} works well for $t\ll \Ms^2$. 

\paragraph{Behavior at large $t$.}
At large $t\gtrsim \Ms^2$, the behavior of $\rho_{st}^A(s,t)$ changes.
This will not be needed in this paper but we record it for future reference. 
In the domain $\Ms^2\ll t\ll s$, it still Regge-behaved and the formula \eqref{eq:type2_regge_estimate1} works, while a prefactor $C(t)$ decays exponentially in $t$, which is simply due to the exponential decay of $f_\VS(t)\sim e^{-\frac{t}{2}\ln(t/4e)}$.
At larger $t\gtrsim s$, $\rho_{st}^A(s,t)$ is no longer given by \eqref{eq:type2_regge_estimate1}. In this domain, we have $\rho_n(s,t)\sim t^{2n+1}$, and hence the maximum $n$ contribution in \eqref{eq:jsum} becomes dominant. In particular, for $s=4n+2$ with $1\ll n\in \mathbb{N}$, we have 
\begin{align}
    \rho_{st}^A(s,t)|_{t\gg s}
    \simeq
    \rho_{n}(s,t)|_{n=\frac{s-2}{4}}
    \simeq 
    \frac{(g m)^2}{64}\frac{(t/4)^{\frac{s}{2}}}{\left[\Gamma\left(\frac{s+2}{4}\right)\right]^2}
    \simeq     \frac{(gm)^2}{128\pi}\frac{1}{e}(et/s)^{\frac{s}{2}}
    \,.
\end{align}

\subsection{Evaluating single-discontinuity from double-discontinuity}\label{sec:Disc_sumrule}

\begin{figure}[tbp]
 \centering
  \includegraphics[width=0.55\textwidth, trim=100 110 120 70,clip]{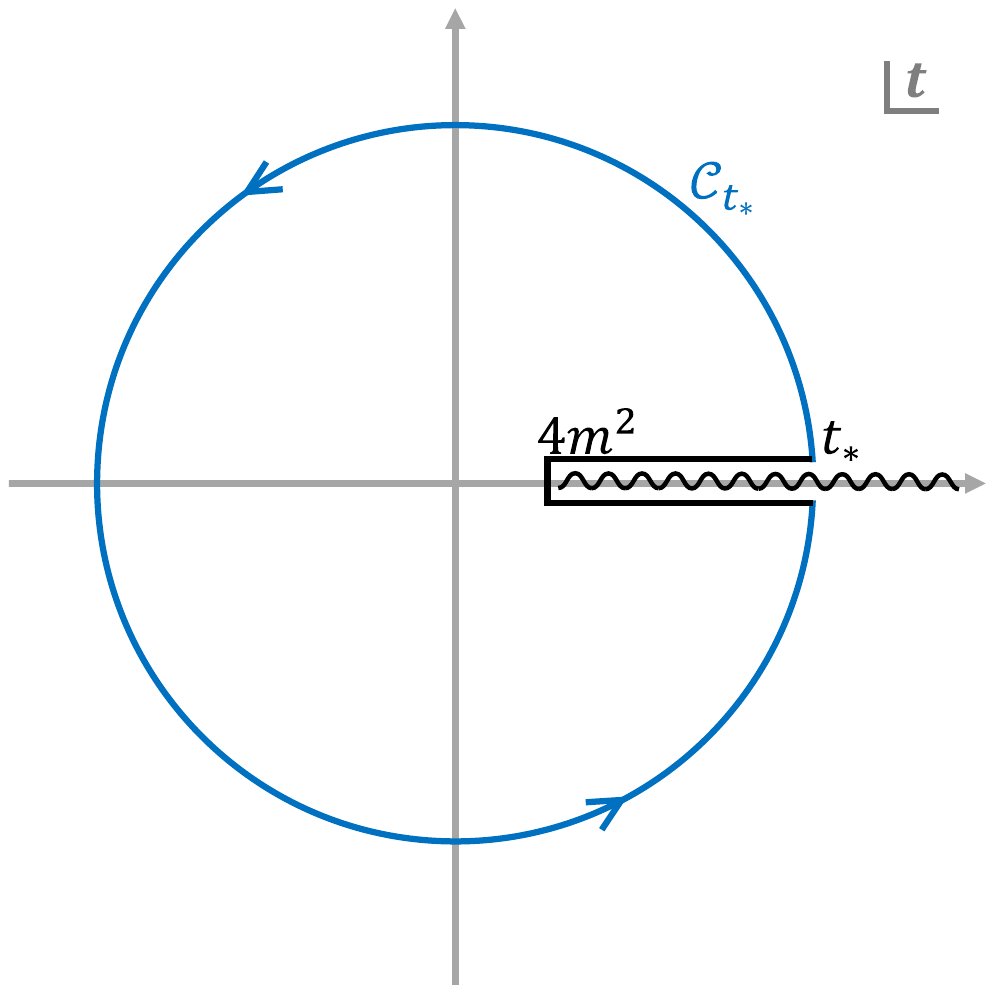}
 \caption{An integration contour in the complex $t$-plane to obtain the formula \eqref{eq:discs_disp} for $\disc_s\scat(s,t)$. In particular, a blue solid line denotes the arc $\mathcal{C}_{t_*}$ with a radius $t_*$ centered at the origin. A branch cut at $t>4m^2$ is due to nonzero $\rho^A_{st}(s,t)$. Note that we have no cut at $4m^2-t_*<t<0$ because we have $\rho_{su}=0$ for the process shown in fig.~\ref{fig:box_closed_1}.}
 \label{fig:contour_t} 
\end{figure}

We now infer the behavior of $\disc_s\scat(s,t)$ given that of $\rho_{st}$.
Formally, we can undo $\disc_t$ by writing a dispersive integral:
\begin{align}
    \disc_s\scat(s,t)
    =
    \int^{t_*}_{4m^2}\frac{\mathrm{d}t'}{\pi}\,
    \frac{\rho_{st}(s,t')}{t'-t}
    +
    \oint_{\mathcal{C}_{t_*}}\frac{\mathrm{d}t'}{2\pi i}\,
    \frac{\disc_s\scat(s,t')}{t'-t}
    \,,\label{eq:discs_disp}
\end{align}
where $\mathcal{C}_{t_*}$ denotes the arc with radius $|t|=t_*$ centered at the origin in the complex $t$-plane, as shown in fig.~\ref{fig:contour_t}.
We choose $m^2\ll t_*\ll \Ms^2$ so that we can make use of the Regge behavior and the above approximations.

While \eqref{eq:discs_disp} is closely related to the idea of the Mandelstam representation, it is important to stress a crucial distinction: we do not integrate up to $t\to\infty$.
In fact, for string amplitudes
$\disc_s\scat(s,t)\sim t^{\alpha(s)}$ at fixed $s$, so such a representation would require an inconveniently large number of subtractions when $\alpha(s)\gg 1$.  This is why we choose instead to close the contour on a finite arc. 
This contour also dispenses us from considering $\rho_{su}$.

In principle, the contribution from the $|t|=t_*$ arc could be evaluated by using $s$-channel unitarity to compute $\disc_s\scat$ using the methods of \cite{Eberhardt:2022zay}.
The crucial point for our purposes is that this calculation cannot yield any $1/m$-enhanced term:
the limit $m\to 0$ is nonsingular since $|t|\gg m^2$ and the calculation reduces to that of a massless process, which is nonsingular.\footnote{Logarithmic divergences can potentially occur at $m\to 0$ but we do not expect power-low divergences in $m$ when fixing the dimensionless coupling $g$ as in section~\ref{sec:review}.}
Thus, since our interest is in $1/m$-enhanced terms, we will simply ignore the arc in \eqref{eq:discs_disp}. 

Eq.~\eqref{eq:type2_regge_estimate1} shows the Reggeization of $\rho_{st}^A(s,t)$. Inserting into \eqref{eq:discs_disp}, this demonstrates the Reggeization of $\disc_s \scat(s,t)$ and therefore also that $\scat(s,t)$, at least as far as $1/m$-enhanced terms are concerned.  Hence, the process shown in fig.~\ref{fig:box_closed_1} is responsible for the Reggeization
of the low-energy triangle diagram of fig.~\ref{fig:ph_dp_grav1} at loop level. 

Let us now describe the extraction of one-loop corrections
$f_\oneloop$ and $\alpha_\oneloop$ from the discontinuity.  Treating these quantities as small, the leading-order corrections to $\disc_s\scat(s,t)$ can be expanded as 
\begin{align}
    \disc_s\scat(s,t)
    =
    \left[
        f(t)+f_\VS(t)\alpha_\oneloop(t)\ln(s/4)+
        \cdots
    \right](s/4)^{\alpha_\tree(t)}
    \,,\label{disc_perturb1}
\end{align}
where the ellipses stand for the higher-order terms.
In the present setup, smallness of corrections are controlled by small coupling constants $g$ and $\Mpl^{-1}$. By taking the discontinuity of \eqref{disc_perturb1} in $t$ and using $\disc_tf_\VS(t>4m^2)=0$, we find that the coefficients of the $\mathcal{O}\bigl((\ln(s/4))^0\bigr)$ term and the $\mathcal{O}\bigl(\ln(s/4)\bigr)$ term of $\rho^\oneloop_{st}$ can be identified as $\disc_tf_\oneloop$ and $\disc_t\alpha_\oneloop$, respectively. Hence, we can evaluate the type-$A$ contributions to $f_\oneloop$ and $\alpha_\oneloop$ from the expression \eqref{eq:type2_regge_estimate1} of $\rho_{st}^A$ as 
\begin{align}
    &\disc_tf^A_\oneloop(t)
    =
    C(t)
    \,,
    \qquad
    \disc_t\alpha^A_\oneloop(t)
    = 0 \,.\label{Im_direct_A}
\end{align}
We find no corrections to the Regge trajectory, which agrees with the observation made in \cite{Hamada:2023cyt}.
We will see how these results are predicted from the $t$-channel partial wave amplitude via the $t$-channel elastic unitarity in subsection~\ref{sec:unitarity}. 

Then, we substitute \eqref{disc_perturb1} into the dispersive formula \eqref{eq:discs_disp} and focus on the term proportional to $(\ln(s/4))^0$ on both sides at the one-loop order, yielding
\begin{align}
    f^A_\oneloop(t)
    &\simeq
    \int^{t_*}_{4m^2}\frac{\mathrm{d}t'}{\pi}\,
    \frac{C(t')}{t'-t}\left(\frac{s}{4}\right)^{\alpha_\tree(t')-\alpha_\tree(t)}
    +
    \oint_{\mathcal{C}_{t_*}}\frac{\mathrm{d}t'}{2\pi i}\,
    \frac{f^A_\oneloop(t')}{t'-t}\left(\frac{s}{4}\right)^{\alpha_\tree(t')-\alpha_\tree(t)}
    \no\\
    &\simeq
    \int^{t_*}_{4m^2}\frac{\mathrm{d}t'}{\pi}\,
    \frac{C(t')}{t'-t}
    +
    \oint_{\mathcal{C}_{t_*}}\frac{\mathrm{d}t'}{2\pi i}\,
    \frac{f^A_\oneloop(t')}{t'-t}
    \qquad (|t|\ll t_*)
    \,,
    \label{eq:discs_disp_A}
\end{align}
where we used $\alpha_\tree(t)\simeq \alpha_\tree(0)$ for $|t|\ll \Ms^2$ in the second line. As explained above, $1/m$-enhanced terms can come only from the first integral.

\subsection{Test of FESR}\label{sec:FESRtest}

\begin{figure}[tbp]
\centering
\begin{fmffile}{cancellation} 
\begin{align*}
\raisebox{-0.45\height}{
\begin{fmfgraph*}(110,70)
\fmfleft{i1,o3}
\fmfright{i2,o4}
\fmf{plain,label.side=left}{v3,o3}
\fmf{plain,label.side=left}{i1,v1}
\fmf{plain,label.side=right}{v2,o4}
\fmf{plain,label.side=right}{i2,v2}
\fmf{plain,
label.side=left,tension=0.7}{v1,v3}
\fmf{plain,
tension=0.35}{v4,v1} 
\fmf{dbl_wiggly,tension=0.7}{v2,v4}
\fmf{plain,
label.side=left,tension=0.35}{v3,v4}
\fmfdot{v1,v3,v2,v4}
\raisebox{-0.15\height}{
\begin{tikzpicture}[overlay]
    \node (n) at (0,0) {};
    \draw [decorate,decoration={brace,amplitude=10pt,mirror,raise=4pt}]
    (0,0) -- (3.6,0) node [black,midway,yshift=-0.8cm] {\footnotesize Graviton exchange in EFT};
\end{tikzpicture}
}
\end{fmfgraph*}
}
\qquad
\longleftrightarrow
\qquad
\raisebox{-0.45\height}{
\begin{fmfgraph*}(110,70)
\fmfleft{i1,m1,o3}
\fmfright{i2,m2,o4}
\fmf{plain,label.side=left}{v3,o3}
\fmf{plain,label.side=left}{i1,v1}
\fmf{plain,label.side=right}{v2,o4}
\fmf{plain,label.side=right}{i2,v2}
\fmf{plain,
label.side=left,tension=0.6}{v1,v3}
\fmf{plain,
tension=0.6}{v4,v1} 
\fmf{dbl_wiggly,label=$g_*$,tension=0}{v2,v4}
\fmf{plain,
label.side=left,tension=0.6}{v3,v4}
\fmfdot{v1,v3,v2,v4}
\fmf{dashes,fore=red}{m1,m2}
\fmf{dashes,fore=red}{v2,v4}
\raisebox{-0.15\height}{
\begin{tikzpicture}[overlay]
    \node (n) at (0,0) {};
    \draw [decorate,decoration={brace,amplitude=10pt,mirror,raise=4pt}]
    (0,0) -- (3.6,0) node [black,midway,yshift=-0.8cm] {\footnotesize Reggeized graviton exchange in the UV theory};
\end{tikzpicture}
}
\end{fmfgraph*}
}
\qquad
\raisebox{-0.45\height}{
\begin{tikzpicture}
    \draw[->,thick] (0,0) -- (0,1.5);
    \node[below] at (0,0) {$s$-channel};
\end{tikzpicture}
}
\end{align*}
\end{fmffile}
\vspace{2mm}
\caption{
Cancellation mechanism for $1/m^2$-enhanced terms in the finite-energy sum rule \eqref{eq:c2_sumrule}. Reggeization of the graviton in the low-energy diagram on the left is naturally related to the high-energy process on the right,
which is singular at small-$t$ for arbitrary high energies.
By channel duality, the latter is equivalent to the sum of box diagrams shown in fig.~\ref{fig:box_closed_1}.
}
\label{fig:mechanism}
\end{figure}
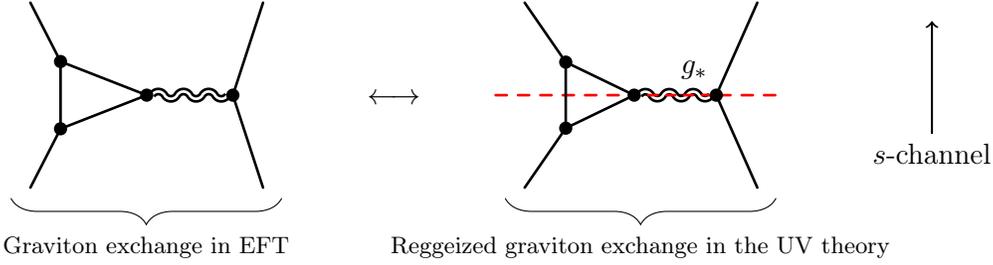

Now we evaluate quantitatively the $m^{-2}$-enhanced contributions from the type-A process to the term $f'/f$ which appears in the first term in the square bracket of \eqref{eq:c2_sumrule}. 
Using \eqref{eq:discs_disp_A} with \eqref{eq:C_app1}, we obtain
\begin{align}
    \der_tf^A_\oneloop(t)|_{t=0}
    &=
    \int^{t_*}_{4m^2}\frac{\mathrm{d}t'}{\pi}\,
    \frac{C(t')}{{t'}^2} 
    +\mathcal{O}(m^0)\no
\\ &= \frac{(gm)^2}{\Mpl^2} \int_{4m^2}^\infty
  \frac{\mathrm{d}t}{2\pi t^2\sqrt{t(t-4m^2)}}
    Q_{2}\left(1+\frac{2{m'}^2}{t-4m^2}\right)
    +\mathcal{O}(m^0)\no
\\ &=
    \frac{g^2}{\Mpl^2}
    \frac{45-8 \sqrt{3} \pi }{1296 \pi  m^2} 
    +\mathcal{O}(m^0)
    \quad \mbox{for $m'=m$}
    \,.
    \label{eq:tdvdisc_1}
\end{align}
The dependence on $t_*$ disappears because the integral is dominated by the lower end $t'\sim 4m^2$, cf. fig.~\ref{fig:ct_single_shape0}.
Using $f(0)\simeq f_\text{VS}(0)= 4\pi/\Mpl^2$,
and multiplying by $2$ to account for the other diagram obtained from fig.~\ref{fig:box_closed_1} by the permutation $(p_1,p_2)\leftrightarrow (p_3,p_4)$,
we predict the contribution to the square bracket of the FESR \eqref{eq:c2_sumrule}:
\begin{align}
    \frac{-2}{\Mpl^2} \frac{\der_tf^A_\oneloop(t)|_{t=0}}{f(0)}
    \times 2
    \simeq
    -\frac{\der_tf^A_\oneloop(t)|_{t=0}}{\pi}
    =
    -\frac{g^2}{\Mpl^2}
    \frac{45-8 \sqrt{3} \pi }{1296 \pi^2  m^2}
    +\mathcal{O}(m^0)
     \,.\label{eq:c2_uvresult}
\end{align}
Eq.~\eqref{eq:c2_uvresult} agrees precisely with the $m^{-2}$-enhanced negative term in $c_2(0)$ predicted by EFT which was recorded in \eqref{c2_eg1}! This is one of the main result of this study, addressing the question raised in section~\ref{sec:review}.
Even though a specific type-II superstring amplitude was used to obtain \eqref{eq:c2_uvresult}, we expect its validity to be more general as discussed 
in subsection~\ref{sec:unitarity}.
The cancellation provides an example of how stringy amplitudes reproduce the prediction of EFT in the low-energy limit and Reggeize the graviton exchange at the loop level. This interesting interplay between low and high-energy processes is summarized in fig.~\ref{fig:mechanism}.

The cancellation has an important implication for gravitational positivity bounds. It highlights the importance of high-energy, near-forward contributions to sum rules, when one takes $t\to 0$ at low energies. After the graviton pole is canceled, this region can leave a large negative remainder enhanced by $m^{-2}$.
In order for the FESR \eqref{eq:c2_sumrule} to constrain EFT spectra, in the way suggested by the Weak Gravity Conjecture for instance, one would need to be able to further constrain the finite remainder and its interplay with the positive first term on the RHS of \eqref{eq:c2_sumrule}. To gain more insights on this aspect, it could be useful to look at other probes of low-energy coefficient, such as the black hole entropy shift~\cite{Cheung:2018cwt} and corrections to the black hole extremality condition~\cite{Hamada:2018dde, Chen:2019qvr,Arkani-Hamed:2021ajd,Noumi:2022ybv,Abe:2023anf}.

It is interesting to connect the negativity observed here with that found using the smeared sum rules of \cite{Caron-Huot:2021rmr}.
The idea there was to reduce IR sensitivity by moving away from the small-$t$
region and exploiting a finite range $-M^2<t<0$.
In the absence of low-energy loops but in the presence of the graviton pole, this reference showed that unitarity allowed some amount of negativity $c_2\gtrsim -\frac{1}{\Mpl^2 M^2}$.
Since this calculation was dominated by a single scale $M$, it seems plausible 
that this tree-level bound will only be affected by light matter loops,
when considering a coupling $c_2'$ measured at the scale $M$ (as opposed to the deep IR), by a small relative amount $\sim\!\frac{g^2}{\Mpl^2 M^2}$, which has a similar parametric form to the above with $m\mapsto M$. This would be interesting to confirm in explicit examples.

\begin{figure}[tbp]
 \centering
  \includegraphics[width=.5\textwidth, trim=0 0 0 0,clip]{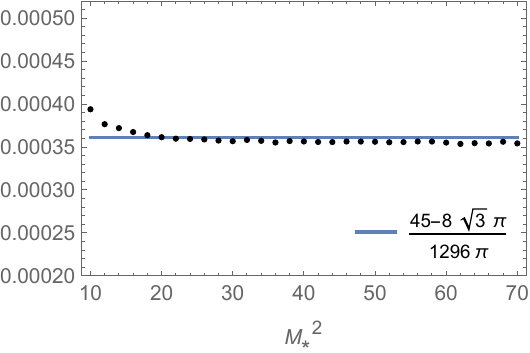}
 \caption{A plot of the RHS of \eqref{eq:FESR0} multiplied by $m^2$ for given $s=10,12,14\cdots 70$. The solid line is the prediction \eqref{eq:c2_uvresult} of the LHS of \eqref{eq:FESR0} multiplied by $m^2$. The figure shows the FESR \eqref{eq:FESR0} well reproduces the result \eqref{eq:c2_uvresult} even for a low value of $s\Mstar^2\gtrsim 15$ for which only contributions from heavy states up to level $n=3$ are contained.}
 \label{fig:FESRtest_tderiv1} 
\end{figure}
Having observed the exact cancellation of the $m^{-2}$-enhanced negative term in the FESR \eqref{eq:c2_sumrule},
it is interesting to ask how low the scale $\Mstar(>\Ms)$
can be taken to be, i.e, how many heavy states are needed to exhibit the channel duality shown in fig.~\ref{fig:box_closed_1}.
Recall that $\Mstar$ denotes the mass scale of Reggeization of graviton exchange.
Here we consider on the FESR for $f(t)$ derived in \cite{Noumi:2022wwf} and ask when it is saturated so that it correctly reproduces the prediction \eqref{eq:tdvdisc_1} for $\der_tf_\oneloop^A(t)|_{t=0}$. When focusing on the $m^{-2}$-enhanced term, the FESR for $\der_tf_\oneloop^A(t)|_{t=0}$ simplifies to
\begin{align}
    \der_tf_\oneloop^A(t)|_{t=0}
    &\approx
    \left(\frac{4}{\Mstar^2}\right)^{2}
    \int^{\Mstar^2}_4\frac{\mathrm{d}s}{s}\,
    y^2(18y^2-8)\int^{t_*}_{4m^2}\frac{\mathrm{d}t}{\pi}\,\frac{\rho_{st}^A(s,t)}{t^2}
    \,,
    \quad
    y\equiv s/\Mstar^2
    \,,\label{eq:FESR0}
\end{align}
and we can numerically evaluate the RHS by using the exact form \eqref{eq:jsum} of $\rho_{st}^A(s,t)$. We describe technical details on the derivation of \eqref{eq:FESR0} and the numerical evaluation in appendix~\ref{sec:FESR}. 
The results are plotted in fig.~\ref{fig:FESRtest_tderiv1}. The figure shows that the FESR \eqref{eq:FESR0} is well satisfied even for a low value of $\Mstar^2\approx 3\Ms^2$ for which only heavy states up to level $n=3$ are taken into account. This is reminiscent of the validity of Dolen-Horn-Schmidt(DHS) duality~\cite{Dolen:1967jr,Dolen:1967zz} for the Veneziano-type amplitudes even at low mass levels~\cite{Ademollo:1967zz,Ademollo:1968cno,Veneziano:1968yb}.

\subsection{Interpretation using \texorpdfstring{$t$}{}-channel partial wave unitarity and generalization}\label{sec:unitarity}
In this subsection, we evaluate the one-loop amplitude $\scat_\oneloop$ involving the type-A process for given tree-level amplitudes $\scat_\tree$ based on the dual channel picture shown in fig.~\ref{fig:box_closed_1}. 
The form of the result \eqref{Cdef_general} suggests that it is convenient to work in the $t$-channel partial wave.
Let us thus start by introducing the $t$-channel partial-wave expansion\footnote{In the presence of $s$ (and/or) $u$-channel massless singularities, this sum only makes sense for $s<0$. However, the elastic unitarity relation which we consider can be analytically continued outside this region.}
\begin{align}
    \scat(s,t)
    =\sum_{J=0}^\infty 16\pi(2J{+}1)f_J(t) P_J\left(1+\frac{2s}{t{-}4m^2}\right)
    \,.
\end{align}
We then introduce the so-called reduced partial wave amplitude $\phi_\ell(t)\equiv ((t-4m^2)/\Ms^2)^{-\ell}f_\ell(t)$, which has a better analytic property in the $t$-plane than $f_\ell$ does for non-integer $\ell$ so that the values of $\phi_\ell(t\pm i\epsilon)$ are the boundary values of the same analytic function.
In terms of $\phi_\ell$, an elastic unitarity condition for $\phi_\ell(t)$ analytically continued to complex $\ell$ is (see e.g.,~\cite{Gribov:2003nw,Correia:2020xtr})
\begin{align}
    &\disc_t\,\phi_\ell(t)
    =
    \Phi_\ell(t)\phi_\ell(t_+)\phi_\ell(t_-)
    \,,\quad
    \Phi_\ell(t)
    \equiv
    \frac{1}{2}\sqrt{\frac{t-4m^2}{t}}
    \left(\frac{t-4m^2}{\Ms^2}\right)^\ell
    \,,
    \label{eq:unitarity}
\end{align}
for $t>4m^2$. Here, $t_\pm\equiv t\pm i\epsilon$. Below, we simply ignore inelastic contributions since in this paper we focus on the two-particle threshold at $t=4m^2$ associated to the exchange of two light particles.

Let us now apply \eqref{eq:unitarity} to our setup. We decompose $\phi_\ell$ into the stringy piece and the EFT piece as $\phi_\ell=\phi_\ell^\text{string}+\phi_\ell^\text{eft}$ with 
\begin{align}
    &\phi_\ell^\text{string}(t)
    = 
    \frac{1}{r(\alpha(t))}
    \frac{f(t)}{\ell-\alpha(t)}
    \,,\quad
    \phi_\ell^\text{eft}(t)
    = 
    \frac{(gm)^2\Ms^{2\ell}}{8\pi(t-4m^2)^{\ell+1}}
    Q_{\ell}\left(1+\frac{2{m'}^2}{t-4m^2}\right)
    +\cdots
    \,,\label{reg}
\end{align}
where the ellipses stand for higher-loop terms, and the constant $r(\alpha)$, related to the asymptotic behavior of Legendre polynomials, is defined by
\begin{align}
    r(\alpha)
    \equiv
    8\pi^2(2\alpha+1)
    \frac{\Gamma(1+2\alpha)}{\left[\Gamma(1+\alpha)\right]^2}
    \,.
\end{align}
The perturbative loop expansions for $\scat_\text{string}$ are parameterized by $f=f_\VS + f_\oneloop+\cdots$ and $\alpha=\alpha_\tree+\alpha_\oneloop+\cdots$. Correspondingly the amplitude is decomposed as $\scat=\scat_\text{string}+\scat_\text{eft}$ with $\disc_s\scat_\text{string}(s,t)\approx f(t)(s/\Ms^2)^{\alpha(t)}$ and $\disc_s\scat_\text{eft}(s,t)=(gm)^2\pi\delta(s-{m'}^2)$. Then $\scat_\text{string}$ and $\scat_\text{eft}$ at the tree-level are the sub-diagrams $\scat_\text{type II}$ and $\scat_\text{scalar}$ of the type-A process, respectively. 

Employing the decomposition of $\phi_\ell$ in \eqref{eq:unitarity}, we find  on the RHS the cross term $\phi_\ell^\text{string}\phi_\ell^\text{eft}$, $(\phi_\ell^\text{string})^2$ term, and the $(\phi_\ell^\text{eft})^2$ term. 
We refer to the contributions to $\disc_t\phi_\ell$ from these terms as A-term, B-term, and C-term, respectively, and then eq.~\eqref{eq:unitarity} gives
\begin{align}
    &\disc_t\phi_\ell(t) 
    = 
    \text{[A-term]}
    +
    \text{[B-term]}
    +
    \text{[C-term]}
    \,,\label{unitarity3}
\end{align}
with
\begin{subequations}
\label{ABCterm}
\begin{align}
    &\text{[A-term]}
    \equiv
    \Phi_\ell(t)
    \left[ 
    \phi^\text{eft}_\ell(t_+) \phi^\text{string}_\ell(t_-)
    +
    \phi^\text{eft}_\ell(t_-) \phi^\text{string}_\ell(t_+)
    \right]
    \,,\label{correct-A}
    \\
    &\text{[B-term]} 
    \equiv \Phi_\ell(t)
    \phi^\text{string}_\ell(t_+)\phi^\text{string}_\ell(t_-)
    \,,\\
    &\text{[C-term]} 
    \equiv \Phi_\ell(t)
    \phi^\text{eft}_\ell(t_+)\phi^\text{eft}_\ell(t_-)
    \,.
\end{align}    
\end{subequations}
The $A$ and $B$-terms can contribute to $\phi_\ell^\text{string}$ while the C-term cannot, since the C-term is regular in the $\ell$-plane. 
Recalling that the correspondence between $(\phi^\text{eft}_\ell, \phi^\text{string}_\ell)$ and the sub-diagrams $(\scat_\text{scalar},\scat_\text{type II})$ of the type-A process, we find that the A-term represents the corrections to the Regge behavior from the type-A process. We therefore solve \eqref{unitarity3} with setting B and C-terms to be zero for a while. Substituting \eqref{reg} into \eqref{correct-A}, we obtain the following solutions at $\mathcal{O}(g^2\Mpl^{-2})$: 
\begin{align}
    &\disc_tf(t)|_\text{A term}
    =
    \frac{f_\VS(t)(gm)^2}{4\pi(t{-}4m^2)^{\alpha_\tree(t)+1}}
    Q_{\alpha_\tree}\left(1{+}\frac{2{m'}^2}{t{-}4m^2}\right)
    \,,
    \quad
    \disc_t\alpha(t)|_\text{A term}=0
    \,.\label{result_a}
\end{align}
This precisely matches with the earlier result \eqref{Im_direct_A}! (In this comparison we include the factor 2 accounting for the permuted diagram obtained from fig.~\ref{fig:box_closed_1} as explained above \eqref{eq:c2_uvresult}.)

Similarly, we can evaluate contributions from the $B$ term to $\disc_t\phi_\ell^\text{string}$ by solving \eqref{unitarity3}: we compute them explicitly in appendix~\ref{sec:reggetheory}, and the results are \eqref{result_b}. Here we simply observe that B term produces a double pole in the $\ell$-plane, which predicts a nontrivial correction to the trajectory
(more precisely, to $\disc_t\alpha$).
Below (in section~\ref{sec:double_regge})
we will confirm this prediction by computing the type-B $s$-channel process shown in fig.~\ref{fig:loopRegge_closed1}.

\paragraph{Generalization.} 

The above discussion based on unitarity holds true for generic form of $f(t)$.
For instance, let us consider the situation where the tree-level term of $f(t)$ is given by some generic function $f_\tree^\text{gen}(t)$ rather than the specific form $f_\VS(t)$ defined in \eqref{eq:fVS}: i.e., we consider the generalized stringy amplitude $\scat_\text{string}$ instead of $\scat_\text{type II}$, which Reggeizes the tree-level graviton exchange as
\begin{align}
    \disc_s\scat_\text{string}(s,t)
    \simeq f_\tree^\text{gen}(t)(s/4)^{\alpha_\tree(t)}
    \quad
    (s\gg\Ms^2,|t|)
    \,.
\end{align} 
Then, the same discussion remains valid and corrections to $f(t)$ and $\alpha(t)$ are simply given by \eqref{result_a} and \eqref{result_b} with $f_\VS(t)\mapsto f_\tree^\text{gen}(t)$.
Accordingly, the double discontinuity $\rho_{st}(s,t)$ at $s\gg\Ms^2,|t|$ for the type-A process is given by \eqref{eq:type2_regge_estimate1} and \eqref{Cdef_general} with $f_\VS\mapsto f_\tree^\text{gen}$, 
and the calculation \eqref{eq:c2_uvresult} of $\der_tf_\oneloop^A/f(0)$ is generalized as
\begin{align}
    \frac{-2}{\Mpl^2} \frac{\der_tf^A_\oneloop(t)|_{t=0}}{f(0)}
    \times 2
    &\simeq
    \frac{-(gm)^2}{\Mpl^2\pi}
    \int^{t_*}_{4m^2}\mathrm{d}t\,
    \frac{Q_{\alpha_\tree(t)}\left(1{+}\frac{2{m'}^2}{t-4m^2}\right)
    }{2\pi t^2\sqrt{t(t{-}4m^2)}}
    \frac{f_\tree^\text{gen}(t)}{f_\tree^\text{gen}(0)}
     \,.\label{eq:c2_uv_general}
\end{align}

\begin{figure}[tbp]
\centering
\begin{fmffile}{regge_general} 
\begin{align*}
\raisebox{-0.45\height}{
\begin{fmfgraph*}(140,70)
\fmfleft{i1,m1,i2}
\fmfright{o1,m2,o2}
\fmf{plain,label.side=left}{i2,v2}
\fmf{plain,label.side=left}{i1,v1}
\fmf{plain,label.side=left}{v3,o2}
\fmf{plain,label.side=left}{v3,o1}
\fmf{plain,label=$m'$,label.side=left,tension=0.3}{v1,v2}
\fmf{plain,label=$m$,label.side=left,tension=0.5}{v2,v3} 
\fmf{plain,label=$m$,label.side=left,tension=0.5}{v3,v1}
\fmfdot{v1,v2,v3}
\fmflabel{$1$}{i1}
\fmflabel{$3$}{i2}
\fmflabel{$2$}{o1}
\fmflabel{$4$}{o2}
\fmfv{decor.shape=circle,decor.filled=5,decor.size=14mm}{v3}
\begin{tikzpicture}[overlay]
    \node at (3.4,1.2) {$\mathcal{M}_\text{string}$};
\end{tikzpicture}
\begin{tikzpicture}[overlay]
     \draw[red, dashed, line width=1pt] (0,1.23) -- (4.7,1.23);
\end{tikzpicture}
\end{fmfgraph*}
}
\qquad
\equiv
\qquad
\raisebox{-0.45\height}{
\begin{fmfgraph*}(110,70)
\fmfleft{i1,m1,o3}
\fmfright{i2,m2,o4}
\fmf{plain,label.side=left}{v3,o3}
\fmf{plain,label.side=left}{i1,v1}
\fmf{plain,label.side=right}{v2,o4}
\fmf{plain,label.side=right}{i2,v2}
\fmf{plain,label=$m'$,label.side=left,tension=0.6}{v1,v3}
\fmf{plain,label=$m$,tension=0.6}{v4,v1} 
\fmf{dbl_wiggly,label=$g_*$,tension=0.1}{v2,v4}
\fmf{plain,label=$m$,label.side=left,tension=0.6}{v3,v4}
\fmfdot{v1,v3,v2,v4}
\fmf{phantom,fore=red}{m1,m2}
\fmf{phantom,fore=red}{v2,v4}
\fmflabel{$1$}{i1}
\fmflabel{$2$}{i2}
\fmflabel{$3$}{o3}
\fmflabel{$4$}{o4}
\begin{tikzpicture}[overlay]
     \draw[red, dashed, line width=1pt] (0,1.23) -- (4,1.23);
\end{tikzpicture}
\end{fmfgraph*}
}
\qquad
\raisebox{-0.45\height}{
\begin{tikzpicture}
    \draw[->,thick] (0,0) -- (0,1.5);
    \node[below] at (0,0) {$s$-channel};
\end{tikzpicture}
}
\end{align*}
\end{fmffile}
\caption{A schematic picture of the channel duality for a generic stringy amplitude $\scat_\text{string}$ which Reggeizes the tree-level graviton exchange. This is a generalization of fig.~\ref{fig:box_closed_1}.
}
\label{fig:Regge_general}
\end{figure}
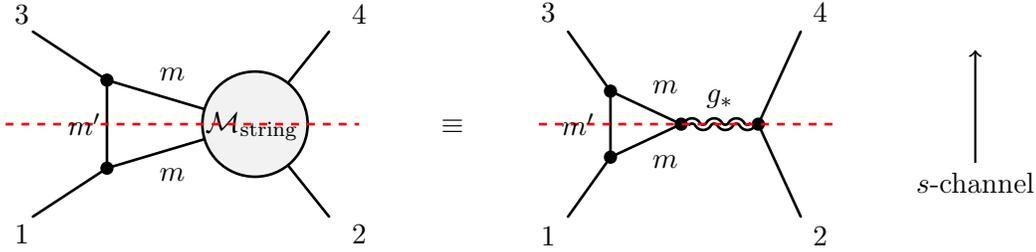

This generalization allows us to include various type of string amplitudes which Reggeize the tree-level graviton exchange, such as the non-planar annulus amplitude for the case of open string scattering (see e.g.,~\cite{Banerjee:2024ibt}). Though the precise behavior of $f_\tree^\text{gen}(t)$ will depend on the details of models or processes, we expect generally that $f_\tree^\text{gen}(t)$ will not possess singularities in the domain $|t|\sim 4m^2\ll\Ms^2$ and can be treated as a constant:
\begin{align}
    f_\tree^\text{gen}(t)
    \simeq 
    f_\tree^\text{gen}(0)
    =
    \frac{\pi\alpha'_\tree\Ms^4}{2\Mpl^2}
    \,,\qquad
    (|t|\ll t_*\,,\quad t_*\gg 4m^2)
    \,.\label{expect}
\end{align}
Here, we use the fact that the value of $f_\tree^\text{gen}(0)/\alpha'_\tree$ is fixed by the residue of the graviton pole, 
which is universal.
When eq.~\eqref{expect} holds, the integral in \eqref{eq:c2_uv_general} reduces to the one evaluated in \eqref{eq:tdvdisc_1}, and eq.~\eqref{eq:c2_uv_general} with $m'=m$ predicts exactly the same $m^{-2}$-enhanced term as \eqref{eq:c2_uvresult} independently of the details of $f_\tree^\text{gen}(t)$ at $t\gg 4m^2$. For this reason, we believe that the exact cancellation observed in section~\ref{sec:FESRtest} 
and the duality shown in fig.~\ref{fig:Regge_general}, which is the generalization of fig.~\ref{fig:box_closed_1}, is robust: we only need the tree-level statement \eqref{expect}.
\footnote{One could also of course consider situations where the mass of light particles in loops are different from the mass of particles in external lines. We expect that the factor $\Phi_\ell(t)$ in \eqref{eq:unitarity} and the precise form of \eqref{eq:formula_doubledisc} will change, without altering the main conclusion.
}
It would still be interesting to explicitly construct $\scat_\text{string}$ in a concrete string setup and confirm the validity of \eqref{expect}, which we leave for future work.

\section{Loop amplitudes involving two Regge towers}
\label{sec:double_regge}
In this section, we consider the type-B process of fig.~\ref{fig:loopRegge_closed1}, which is the box-type diagram with $s$-channel exchange of two stringy towers.
We will see that this process produces a threshold singularity of the Regge trajectory $\alpha(t)$ at $t<4m^2$.
In section~\ref{sec:open_model1} we calculate double discontinuity. In section~\ref{sec:trajectory} we calculate corrections to Regge trajectory and find agreement with the results obtained above by solving the analytically-continued $t$-channel partial wave unitarity. 
We also briefly discuss the generalization to generic Regge trajectories.

\subsection{Exchange of two degenerate trajectories}\label{sec:open_model1}
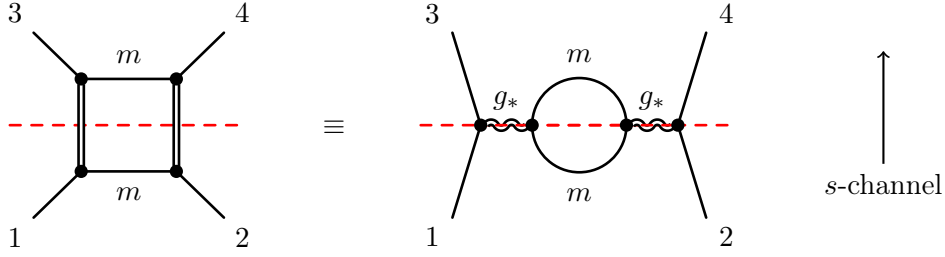
\begin{figure}[tbp]
\centering
\begin{fmffile}{type_B} 
\begin{align*}
\raisebox{-0.45\height}{
\begin{fmfgraph*}(90,70)
\fmfleft{i1,m1,i2}
\fmfright{o1,m2,o2}
\fmf{plain,label.side=left}{i2,v2}
\fmf{plain,label.side=left}{i1,v1}
\fmf{plain,label.side=left}{v4,o2}
\fmf{plain,label.side=left}{v3,o1}
\fmf{dbl_plain,tension=0.5}{v1,v2}
\fmf{plain,label=$m$,tension=0.5}{v4,v2} 
\fmf{dbl_plain,tension=0.5}{v3,v4}
\fmf{plain,label=$m$,tension=0.5}{v1,v3}
\fmfdot{v1,v2,v3,v4}
\fmflabel{$1$}{i1}
\fmflabel{$3$}{i2}
\fmflabel{$2$}{o1}
\fmflabel{$4$}{o2}
\fmf{dashes,fore=red}{m1,m2}
\end{fmfgraph*}
}
\qquad
\equiv
\qquad
\raisebox{-0.45\height}{
\begin{fmfgraph*}(120,70)
\fmfleft{i1,m1,o3}
\fmfright{i2,m2,o4}
\fmf{plain}{v1,o3}
\fmf{plain}{v1,i1}
\fmf{plain}{v2,o4}
\fmf{plain}{v2,i2}
\fmf{dbl_wiggly,label=$g_*$,tension=0.1}{v3,v1}
\fmf{plain,left,label=$m$,tension=-0.2}{v3,v4,v3}
\fmf{dbl_wiggly,label=$g_*$,tension=0.1}{v2,v4}
\fmfdot{v1,v3,v2,v4}
\fmf{dashes,fore=red}{m1,m2}
\fmf{dashes,fore=red}{v2,v4}
\fmf{dashes,fore=red}{v4,v3}
\fmf{dashes,fore=red}{v3,v1}
\fmflabel{$1$}{i1}
\fmflabel{$2$}{i2}
\fmflabel{$3$}{o3}
\fmflabel{$4$}{o4}
\end{fmfgraph*}
}
\qquad
\raisebox{-0.45\height}{
\begin{tikzpicture}
    \draw[->,thick] (0,0) -- (0,1.5);
    \node[below] at (0,0) {$s$-channel};
\end{tikzpicture}
}
\end{align*}
\end{fmffile}
\caption{A channel duality for $\disc_s\scat_B$ for the type-B process. Solid lines represent a massive scalars with mass $m$. The mass for internal lines is written explicitly for definiteness. A solid double line represents the sum of whole heavy states on the Regge trajectory. 
}
\label{fig:dual_B} 
\end{figure}

We consider the diagram shown in fig.~\ref{fig:dual_B} for which the $t$-channel discontinuity of the amplitude is given by the product of same amplitude which is Reggeized at $s\gg \Ms^2, t$. Then the double discontinuity $\rho_{st}^B\equiv\disc_t\disc_s\scat_B$ is calculated by using \eqref{eq:formula_doubledisc} with $\disc_{s}\scat_L=\disc_{s}\scat_R$. We parameterize the Regge amplitude at large $s$ as   
\begin{align}
    \disc_{s}\scat_L (s,t)
    =\disc_s\scat_R(s,t)
    \approx
    f_\VS(t)(s/4)^{\alpha_\tree(t)}
    \,,\label{eq:regge_approx}
\end{align}
where sub-leading terms which do not grow as fast as $s^{\alpha_\tree}$ are suppressed.
Below, we focus on the regime $t \ll \Ms^2$ and sufficiently large $s$ satisfying $(t-4m^2)s \gg \Ms^4$ so that we have nontrivial contributions from the configurations $s_Ls_R \geq \Ms^2$.  
We substitute the Regge behavior \eqref{eq:regge_approx} into \eqref{eq:formula_doubledisc} and obtain
\begin{align}
    \rho_{st}^B(s,t)
    \approx
    &\frac{f_\VS^2(t)}{8\pi^2\sqrt{t(t-4m^2)}}
\int^{\bar s}_4\mathrm{d}s_L
    \int^{s_-(s,t;s_L)}_{4}\mathrm{d}s_R\,
    \frac{(s_L/4)^{\alpha_\tree(t)}(s_R/4)^{\alpha_\tree(t)}}{(s_R-s_-)^{1/2}(s_R-s_+)^{1/2}}
    \,,\label{rho_rr_asy1}
\end{align}
where we define $\bar s$ as a solution of $s_-(s,t;\bar s)=4$: we have $\bar s \simeq (t-4m^2)s/16$ by using an approximation \eqref{eq:range_est2} for $s_-$. 
Strictly speaking, the integral in the domain $s_{L/R}\sim 4$ may not be well approximated by using the Regge amplitude: the sub-leading terms suppressed in \eqref{eq:regge_approx} would be important as well. However, this error does not affect the leading-order terms of $\rho^B_{st}(s,t)$ (and hence the $t$-dependence of $\disc_s\scat_B$) at large $s$. This is because,
as we see below, the dominant contribution of \eqref{rho_rr_asy1} comes from the integral in the domain $4 \ll s_{L/R}\ll\bar s$. That is, the summation over many heavy states is essential. This also justifies to 
replace the lower end of $\int\mathrm{d}s_R$ in \eqref{rho_rr_asy1} by $0$. Using this, we can perform $\int\mathrm{d}s_R$ in \eqref{rho_rr_asy1} by using the formula \eqref{int_formula} (which was also employed to get \eqref{eq:type2_regge_estimate1}) as
\begin{align}
    \rho_{st}^B(s,t)
    \approx
    &\frac{f_\VS^2(t)\Gamma(1+\alpha_\tree(t))}{8\pi^{3/2}\sqrt{t(t-4m^2)}\Gamma(3/2+\alpha_\tree(t))}
    \int^{\bar s}_4\mathrm{d}s_L\,(s_L/4)^{\alpha_\tree(t)} 
    \no\\
    &\times
    \left(\frac{s_-}{4}\right)^{\alpha_\tree(t)}
    \sqrt{\frac{s_-}{s_+-s_-}} {}_2F_1\left(\frac{1}{2},\frac{1}{2};\frac{3}{2}+\alpha_\tree(t);-\frac{s_-}{s_+-s_-}\right)
    \,.\label{rho_rr_asy_int}
\end{align}
In the domain of integration, we can use $0<\frac{s_-}{(s_+-s_-)}\simeq (\frac{t-4m^2}{4s_L})^2\ll1$ which follows from \eqref{eq:range_est}, and the approximation ${}_2F_1\left(\frac{1}{2},\frac{1}{2};\frac{3}{2}+\alpha_\tree(t);-\frac{s_-}{s_+-s_-}\right)\simeq 1$. Using these approximations, we can calculate the integral in \eqref{rho_rr_asy_int} as
\begin{align}
    &\int^{\bar s}_4\mathrm{d}s_L\, \left(\frac{s_L}{4}\right)^{\alpha_\tree(t)}
    \left(\frac{s_-}{4}\right)^{\alpha_\tree(t)}
    \sqrt{\frac{s_-}{s_+-s_-}} {}_2F_1\left(\frac{1}{2},\frac{1}{2};\frac{3}{2}+\alpha_\tree(t);-\frac{s_-}{s_+-s_-}\right)
    \no\\
    &
    \simeq    
    \int^{\bar s}_4\mathrm{d}s_L\,
    \left(\frac{s_L s}{16}\right)^{\alpha_\tree(t)}
    \left(\frac{t-4m^2}{4 s_L}\right)^{\alpha_\tree(t)+1}
    =
    \frac{(t-4m^2)^{\alpha_\tree(t)+1}}{4^{2\alpha_\tree(t)+1}} 
    \left(\frac{s}{4}\right)^{\alpha_\tree(t)}
    \ln(\bar s/4)
    \,.\label{eval_RR}
\end{align}
From the definition of $\bar s$, we have $\ln(\bar s/4)\simeq \ln(s/4)+\ln((t-4m^2)/16)$. Thus, we have 
\begin{align}
    \rho_{st}^B(s,t)
    &\approx
    \frac{f_\VS^2(t)\Gamma(1{+}\alpha_\tree(t))(t{-}4m^2)^{\alpha_\tree(t)+\frac{1}{2}}}{2^{4\alpha_\tree(t)+5}\pi^\frac{3}{2}\Gamma\left(\frac{3}{2}{+}\alpha_\tree(t)\right)\sqrt{t}}
    \left(\frac{s}{4}\right)^{\alpha_\tree(t)}
    \left[ 
        \ln(s/4) + \ln(t{-}4m^2) 
    \right]\Theta(t{-}4m^2)
    \label{rho_rr_asy2}
\end{align}
up to $\mathcal{O}(s^{\alpha_\tree(t)}(\ln s)^0)$ terms which stem from the errors induced by  
the Regge amplitude approximation \eqref{eq:regge_approx} and the approximation we used to obtain \eqref{rho_rr_asy_int}.  
Nonetheless, we keep the second $\ln(t-4m^2)$ term since it comes from the expression of $\bar s$ and hence it universally appears irrespective of other sub-leading terms. 

\subsection{Corrections to graviton Regge trajectory}\label{sec:trajectory}
Let us now infer the corrections to the Regge parameters $f(t)$ and $\alpha(t)$ from the type-$B$ process. 
As explained below \eqref{disc_perturb1}, we can evaluate the type-B contributions to $f_\oneloop$ and $\alpha_\oneloop$ from the expression \eqref{rho_rr_asy2} of $\rho^B_{st}$ as
\begin{subequations}
\label{eq:Im_direct_B}
\begin{align}
    &\disc_t\alpha_B(t)
    =
    \frac{f_\VS(t)\Gamma(1{+}\alpha_\tree(t))(t-4m^2)^{\alpha_\tree(t)+\frac{1}{2}}}{\pi^{\frac{3}{2}}2^{4\alpha_\tree(t)+7}\Gamma(\frac{3}{2}{+}\alpha_\tree(t))\sqrt{t}}\Theta(t-4m^2)
    \,,\label{eq:Imalpha_direct_B}
    \\
    &\disc_tf_B(t)
    =
    f_\VS(t)\disc_t\alpha_B(t)
    \ln(t-4m^2) + \cdots
    \,,\label{eq:Imf_direct_B}
\end{align}
\end{subequations}
where the ellipses stand for the contributions from sub-leading terms suppressed in \eqref{rho_rr_asy2}.
In contrast to the type-A process discussed in the previous section, we find corrections to the Regge trajectory.

We see that these results match with the solution \eqref{result_b} of the analytically-continued $t$-channel partial wave unitarity \eqref{unitarity3} with keeping only the B term (Regge-pole term), confirming the overall consistency of the picture developed so far.

From \eqref{eq:Im_direct_B} we can learn how the $t$-dependence of $\alpha(t)$ is corrected by the type-B process. We start with the formula \eqref{eq:discs_disp}. 
We substitute \eqref{disc_perturb1} into the dispersive formula \eqref{eq:discs_disp} and focus on the terms proportional to $\ln(s/4)$ on both sides at the one-loop order, we obtain
\begin{align}
    \alpha_B(t)
    &\simeq
    \int^{t_*}_{4m^2}\frac{\mathrm{d}t'}{\pi}\,
    \frac{\disc_{t'}\alpha_B(t')}{t'-t}
    \frac{f_\VS(t')(s/4)^{\alpha_\tree(t')}}{f_\VS(t)(s/4)^{\alpha_\tree(t)}}
    +
    \oint_{|t'|=t_*}\frac{\mathrm{d}t'}{2\pi i}\,
    \frac{\alpha_B(t')}{t'-t}
    \frac{f_\VS(t')(s/4)^{\alpha_\tree(t')}}{f_\VS(t)(s/4)^{\alpha_\tree(t)}}
    \no\\
    &\simeq
    \int^{t_*}_{4m^2}\frac{\mathrm{d}t'}{\pi}\,
    \frac{\disc_{t'}\alpha_B(t')}{t'-t}
    +
    \oint_{|t'|=t_*}\frac{\mathrm{d}t'}{2\pi i}\,
    \frac{\alpha_B(t')}{t'-t}
    \qquad
    (|t|\ll t_*)
    \,,\label{eq:alphaB_disp}
\end{align}
where we used $f_\VS(t)\simeq f_\VS(0)$ and $\alpha_\tree(t)\simeq \alpha_\tree(0)$ for $|t|\ll1$ in the second line. As explained in section~\ref{sec:Disc_sumrule}, the dominant $t$-dependence in the domain $|t|\ll t_*$ can only come from the first integral.
However, from \eqref{eq:Imalpha_direct_B} we find
\begin{align}
    \der_t^n
    \left[
    \int^{t_*}_{4m^2}\frac{\mathrm{d}t'}{\pi}\,
    \frac{\disc_{t'}\alpha_B(t')}{t'-t}
    \right]_{t=0}
    &\sim
    \int^{t_*}_{4m^2}\frac{\mathrm{d}t'}{\pi}\,
    \frac{(t'-4m^2)^{5/2}}{t'^{n+3/2}}
    =
    \int^{\frac{t_*}{4m^2}-1}_0\mathrm{d}\sigma_{t'}\,
    \frac{(4m^2)^{2-n}\sigma_{t'}^{5/2}}{(\sigma_{t'}+1)^{n+3/2}}
    \,,\label{eq:alphaB_deriv}
\end{align}
which is not dominated by the IR domain $t\sim\mathcal{O}(m^2)$ for the $n=0,1,2$ case. This means that $\der_t^{n}\alpha(t)|_{t=0}$ with $n\leq 2$ as well as the last term $\alpha''/\alpha'$ on the RHS of sum rule \eqref{eq:c2_sumrule} are not enhanced by $1/m$.\footnote{It may be interesting that the $n=2$ case of \eqref{eq:alphaB_deriv} shows a logarithmic divergence in the $m\to 0$ limit, though such a term may also arise from the second term of \eqref{eq:alphaB_disp}.}  
On the other hand, $\der_t^n\alpha(t)|_{t=0}$ with $n\geq 3$ cases receive $1/m$-enhanced corrections from the first term of \eqref{eq:alphaB_disp}. Hence, we can extract the genuinely IR-sensitive component of $\alpha_B(t)$ by rewriting \eqref{eq:alphaB_disp} as
\begin{align}
    \alpha_B(t)
    &\simeq
    \alpha_B(0)
    +\alpha_B't + \frac{\alpha_B''t^2}{2}
    +
    \alpha
    t^3\int^{t_*}_{4m^2}\frac{\mathrm{d}t'}{\pi}\,
    \frac{\disc_{t'}\alpha_B(t')}{{t'}^3(t'-t)}
    \qquad
    (|t|\ll t_*) 
    \,,\label{eq:alphaB_subtdisp}
\end{align}
where the first three terms are not enhanced by powers of $1/m$. 
Eq.~\eqref{eq:alphaB_subtdisp} is a reminiscent of a dispersion relation with subtractions. 
We however emphasize that our derivation of \eqref{eq:alphaB_disp} and \eqref{eq:alphaB_subtdisp} simply relies on the  formula \eqref{eq:discs_disp} for $\disc_s\scat(s,t)$.

Since the type-B process produces the lightest singularity in $\alpha(t)$, we expect that the final term of \eqref{eq:alphaB_subtdisp} governs the leading-order $t$-dependence of graviton Regge trajectory $\alpha(t)$ at $|t|\ll t_*$ with $4m^2\ll t_*\ll1$: 
\begin{align}
    &\alpha(t)
    \simeq
    2 + 
    \alpha't + \frac{\alpha''t^2}{2}
    + \frac{1}{\Mpl^2} h(t)
    \qquad
    (|t|\ll t_*)
    \,,\label{gravtraj}
\end{align}
where $h(t)$ is the final term of \eqref{eq:alphaB_subtdisp} which is evaluated by using the approximate expression of $\disc_t\alpha_B(t)$ at $t\ll 1$ and $f_\VS(0)=4\pi/\Mpl^2$ as
\begin{align}
    &h(t)=
    \frac{240 m^4{-}140 m^2 t{+}23 t^2}{57600 \pi ^2}-\frac{\left(4 m^2{-}t\right)^\frac{5}{2} \log \left(\frac{\sqrt{4 m^2{-}t}+\sqrt{{-}t}}{2 m}\right)}{3840 \pi ^2 \sqrt{{-}t}}
    =\frac{t^3}{107520 \pi ^2 m^2} + \mathcal{O}(t^4)
    \,.\label{def_h}
\end{align}
$h(t)$ is independent of $t_*$ since the integral is dominated by the contributions from the domain $t\ll t_*$. We arrive at the formula eq.~\eqref{gravtraj} with \eqref{def_h}, which captures IR-sensitive component of $\alpha(t)$. Our method does not fix a few parameters $\alpha'$ and $\alpha''$ which are not enhanced by $1/m$. Similarly to the case of type-A process, the result \eqref{gravtraj} can be extended to more general cases when \eqref{expect} is satisfied.

The absence of IR enhancement in $\alpha'(0)$ and $\alpha''(0)$ for the graviton Regge trajectory is due to the behavior $\disc_t\alpha_B\sim (t-4m^2)^{\alpha_\tree(t)}\sim (t-4m^2)^2$ at $t\sim \mathcal{O}(m^2)$.
For a generic trajectory with smaller intercept, the trajectory becomes more IR sensitive: e.g., for a Reggeized photon $\alpha(4m^2)\simeq \alpha(0)=1$ case, $\alpha''$ can be also enhanced by powers of $1/m$. A qualitative interpretation of the factor $(t-4m^2)^{\alpha_\tree(t)}$ is that the interaction vertices in the $t$-channel diagram on the right of fig.~\ref{fig:dual_B} contain more derivatives when the exchanged particles have a higher spin, and some of these derivatives contract to give $t$.

\section{Conclusion}\label{sec:conclusion}

We studied string loop corrections to positivity bounds on low-energy gravitational EFTs in the presence of light massive particles with mass $m\ll\Ms$, in a scenario where the graviton displays Regge behavior in some energy range $\Ms^2\lesssim s \ll \Mpl^2$.
This is a generic situation in weakly coupled string theory.
A peculiar feature of this setup is the presence of ``wrong-sign'' contributions from the regime of high-energy, near-forward scattering, $|t|\sim m^2\ll \Ms^2\lesssim |s|$. The sign  
has interesting implications in the contexts of the Weak Gravity Conjecture and phenomenology, as reviewed in section \ref{sec:review}.

The origin of negative contributions
can be understood intuitively from a simple model. Assuming Regge behavior and a smooth forward limit of the imaginary part at high energies for the purposes of this explanation, the dispersive integral takes the schematic form (with
$M_s^{-2}=\alpha'=1$ for illustration):
\begin{align} \label{eq:schematic sign}
    \int_1^\infty \frac{\mathrm{d}s}{s^3} \disc_s\scat(s,t) 
    &=
    \int_1^\infty \frac{ds}{s^3} s^{2{+}t{+}\gamma t^2{+}O(t^3)}
    (1{+}\beta t+O(t^2)) 
   = \frac{1}{{-}t} - \beta + \gamma+O(t)
    \quad (t<0)
    \,,
\end{align}
where $\beta\propto f'/f$ and $\gamma \propto \alpha''/\alpha'$ in the notation of this paper.
Even though $\beta>0$ by unitarity, it is striking that its contribution after integrating and subtracting the graviton pole yields a finite remainder $-\beta$ that is negative definite.
The usual argument that the coefficient of $s^2t^0$ is a positive sum of partial waves is invalidated by the divergence of the sum/integral at $t=0$.  More robust sum rules, such as the finite-energy sum rule (FESR) \eqref{eq:c2_sumrule} (in which an arc contribution is added to remove sensitivity to very high energies) exhibit the same phenomenon. 
The terms corresponding to $-\beta+\gamma$ in the FESR are given in the square bracket of \eqref{eq:c2_sumrule}. 

The main observation of this paper is that the contribution $-\beta$ not only can be negative, it can also be large.  That is, even though it originates from high-energy processes, it can receive loop corrections that are $1/m$-enhanced in the presence of IR scales. We demonstrated this
by developing a method to precisely evaluate the $t$-dependence of Regge trajectories $\alpha(t)$ and couplings $f(t)$ in the IR domain $|t|\lesssim \mathcal{O}(m^2)$ induced by one-loop corrections, for a given string amplitude which Reggeizes the tree-level graviton exchange.
Technically, starting from the rigorous $s$-channel unitarity formula, we analyzed the double-discontinuity $\rho_{st}$ at one-loop and then ``un-did'' the $t$-channel discontinuity. This suffices to precisely evaluate $1/m$-enhanced terms on the RHS in \eqref{eq:c2_sumrule}. We found that only the coupling $f(t)$ is subject to the large IR corrections at $\mathcal{O}(g^2\Mpl^{-2}m^{-2})$. The negative contribution from 
$-f'/f$ then exactly matches and cancels the $1/m$-enhanced negative term \eqref{c2_eg1} (analogous to \eqref{c2_eg1_intro} for photon-photon scattering) predicted by EFT calculations, ensuring that the FESR is satisfied.

This provides an example of an interesting interplay between low and high-energy processes in string amplitudes.  The mechanism is in fact quite straightforward and summarized in fig.~\ref{fig:mechanism}: it amounts to inserting in a loop diagram the tree-level Reggeization of the graviton propagator.
Note that the possibility of having such large negative terms has been raised in earlier works, see for example~\cite{Caron-Huot:2021rmr,Alberte:2021dnj,Noumi:2022wwf,deRham:2022gfe,Hamada:2023cyt}. Here we demonstrated a simple mechanism that makes such terms large and negative, and evaluated them precisely in calculable models.

Our calculations were performed in the specific string-inspired model described in section~\ref{sec:problem}.
We expect our results to hold quite generally, however,
as further discussed in section~\ref{sec:unitarity}.
Intuitively, $1/m$-enhanced terms can only originate from scattering processes at large impact parameters $b\sim m^{-1}\gg \Ms^{-1}$; they thus necessarily involve the exchange of nearly on-shell light particles. This is why we expect these contributions to be reliably calculable from an
(analytic continuation of) $t$-channel unitarity, that is, 
by multiplying $t$-channel factorized on-shell amplitudes.
As long as tree-level $t$-channel partial waves possess the
smoothness properties in $t$ and $\ell$ near the graviton Regge pole that we assumed,
our results will be technically unchanged.
It would be interesting to confirm that explicit string compactifications indeed lead to scattering amplitudes that satisfy these properties, interpolating between 
$\scat_{\rm EFT}$ at low energies and Regge behavior at high energies. 

It would also be interesting to analyze in similar string scenarios the contribution of heavy
states to other probes of low-energy coefficients, such as 
the black hole entropy shift~\cite{Cheung:2018cwt}.
Clarifying the relationship between the sign of individual contributions and that of infinite sums could be particularly valuable. 

For clarity, we focused on situations where the Weak Gravity Conjecture is parametrically satisfied: $g\gg m/\Mpl$ (analogous to $e\gg m_e/\Mpl$ in QED). 
However, even if one reduces the coupling $g$ so that purely gravitational loop contributions of order $1/\Mpl^4$ become no longer negligible, the source of negativity which we identified will also continue to have a comparable size to these loop contributions, making it nontrivial to predict the sign of the overall sum. 
The same mechanism could potentially affect naive unitarity-based conclusions about the sign of other loop effects, such as the renormalization group running of EFT couplings in the deep infrared.

There exists natural ways to shield oneself from such enhanced negative terms by focusing on small impact parameters \cite{Camanho:2014apa} or by using values of $t$ away from the forward limit \cite{Caron-Huot:2021rmr}, however these also naturally lead to bounds that are less sensitive to the masses of light states.
In order for the FESR to constrain light states, it seems that one would have to constrain the interplay between the positive first term on the RHS of \eqref{eq:c2_sumrule} and the $1/m$-enhanced negative term we discussed. 
It remains an unsolved question whether robust unitarity-based arguments can yield interesting bounds on EFT spectra, such as the masses or relevant interactions
of weakly coupled light particles.

\acknowledgments
We would like to thank Sebastian Mizera and Toshifumi Noumi for useful discussions. J.T. would like to thank the members of McGill university for their hospitality, during which this work was initiated. Work of S.C.H. is supported in parts by the National Science and Engineering Council of Canada (NSERC), the Canada Research Chair program, reference number CRC-2022-00421,
and by the Simons Collaboration on the Nonperturbative Bootstrap.
Work of J.T. is supported in parts by IBS under project code IBS-R018-D1.


\appendix

\section{A formula for calculating double discontinuity}\label{sec:doubledisc_derivation}
In this section, we briefly outline a derivation of the useful formula \eqref{eq:formula_doubledisc} to evaluate the double discontinuity, following \cite{Correia:2020xtr}. We refer to \cite{Correia:2020xtr} for more details. 

Let us consider the scattering amplitude $\scat(s,t)$ of identical massive scalar with mass $m$ in spacetime dimensions $d\geq 4$. We suppose that the $t$-channel discontinuity of the amplitude is given by the product of amplitudes $\scat_L$ and $\scat_R$ as represented in fig.~\ref{fig:general_box}, where these are the scattering amplitudes of scalars with identical mass $m$:
\begin{align}
    \disc_t \scat(s,t)
    = f_d(t)
    \oint_{[-1,1]}\frac{\mathrm{d}\eta_L}{2\pi i}\oint_{[-1,1]}\frac{\mathrm{d}\eta_R}{2\pi i}
    \,
    \widetilde\scat_L(\eta_L,t) \widetilde\scat^*_R(\eta_R,t)
    \times K_d(z_s,\eta_R,\eta_L)
    \,,\label{eq:tdisc_type2}
\end{align}
where $[-1,1]$ denotes a counterclockwise contour around the interval $-1<\eta_{L/R}<1$, and $\eta_L$, $\eta_R$, and $z_s$ are angle variables: $\eta_{L/R}=1+\frac{2s_{L/R}}{t-4m^2}$ and $z_s=1+\frac{2s}{t-4m^2}$.  We define $\widetilde\scat$ as $\widetilde\scat(z_s,t)\equiv\scat(s,t)$. A prefactor $f_d(t)$ and the Mandelstam Kernel $K_d(z,\eta_L,\eta_R)$ are given by
\begin{align}
    &f_d(t)
    \equiv
    \frac{(t-4m^2)^{\frac{d-3}{2}}}{8(4\pi)^{d-2}\sqrt{t}}
    \,,\quad
    K_d(z,\eta_L,\eta_R)
    \equiv
    \int^1_{-1}\mathrm{d}z'\int^1_{-1}\mathrm{d}z''
    \frac{\mathcal P_d(z,z',z'')}{(\eta_L-z')(\eta_R-z'')}
    \,,
\end{align}
where 
\begin{align}
    \mathcal P_{d\geq 4}(z,z',z'')
    \equiv
    \frac{2\pi^{\frac{d-3}{2}}}{\Gamma\left(\frac{d-3}{2}\right)}
    (1-z^2)^{\frac{4-d}{2}}
    \frac{\Theta\left(1-z^2-{z'}^2-{z''}^2+2zz'z''\right)}{\left(1-z^2-{z'}^2-{z''}^2+2zz'z''\right)^{\frac{5-d}{2}}}
    \,.
\end{align}

Eq.~\eqref{eq:tdisc_type2} can be analytically continued to complex $z_s$ thanks to the nice analyticity property of the Mandelstam kernel $K_d$. Hence, \eqref{eq:tdisc_type2} is suitble for evaluating the $st$-double discontinuity $\rho_{st}(s,t)$ in the regime $4m^2\leq t<16m^2$ with $s>0$. To compute it, we only use the discontinuity of $K_d(z,\eta_L,\eta_R)$ in the region $z>1$, $\eta_L\eta_R>0$:
\begin{align}
    &\disc_z K_{d\geq 4}(z,\eta_L,\eta_R)
    =
    \frac{4\pi^\frac{d+1}{2}}{\Gamma\left(\frac{d-3}{2}\right)}
    \Theta(z-\eta_+)
    \frac{(z^2-1)^{\frac{4-d}{2}}}{(z-\eta_-)^{\frac{5-d}{2}}(z-\eta_+)^{\frac{5-d}{2}}}
    \,,
\end{align}
where $\eta_\pm$ are functions of $\eta_L$ and $\eta_R$ defined by 
\begin{align}
    \eta_\pm(\eta_L,\eta_R)
    =
    \eta_L\eta_R
    \pm
    \sqrt{\eta_L^2-1}\sqrt{\eta_R^2-1}
    \,.\label{eq:def_etapm}
\end{align}
Let us now set $d=4$.
The $st$-double discontinuity $\rho_{st}(s,t)$ is 
\begin{align}
    \rho_{st}(s,t)
&    =
    f_4(t)
    \oint_{[-1,1]}\frac{\mathrm{d}\eta_L}{2\pi i}\oint_{[-1,1]}\frac{\mathrm{d}\eta_R}{2\pi i}
    \,
    \widetilde\scat_L(\eta_L,t) \widetilde\scat^*_R(\eta_R,t)
    \times \disc_z K_4(z_s,\eta_R,\eta_L)
    \,.
\end{align}
Deforming the integration contours and substituting the explicit form of $f_4(t)$ and $\disc_zK_4$ into the equation, we obtain 
\begin{align}    
   \rho_{st}(s,t)
&   =\frac{1}{16\pi^2}\sqrt{\frac{t-4m^2}{t}}
    \no\\
&    \qquad\times
\int^\infty_{1}\mathrm{d}\eta_L\int^\infty_{1}\mathrm{d}\eta_R
    \frac{\disc_{\eta_L}\widetilde\scat_L(\eta_L,t)
    \disc_{\eta_R}\widetilde\scat^*_R(\eta_R,t)}{(z_s-\eta_-)^{1/2}(z_s-\eta_+)^{1/2}}\Theta(z_s-\eta_+)
    \,.
\end{align}
We rewrite the integration variables in terms of $s_L$ and $s_R$ and obtain eq.~\eqref{eq:formula_doubledisc}.

\section{Derivation of \texorpdfstring{\eqref{eq:type2_regge_estimate1} and \eqref{Cdef_general}}{a}}\label{sec:compute_C}
Here we derive eqs.~\eqref{eq:type2_regge_estimate1} and \eqref{Cdef_general}. We can perform the integral in \eqref{eq:type2_regge_estimate1} exactly by using the following formula
\begin{align}
    &\int^a_0\mathrm{d}s\,
    \frac{s^c}{(s-a)^{1/2}(s-b)^{1/2}}
    =
    \frac{\sqrt{\pi } a^{c+1} \Gamma (c+1) \, _2F_1\left(\frac{1}{2},\frac{1}{2};c+\frac{3}{2};\frac{a}{a-b}\right)}{\sqrt{a (b-a)} \Gamma \left(c+\frac{3}{2}\right)}
    \quad
    (0<a<b)
    \,.\label{int_formula}
\end{align}
This leads to the following expression for $C(t)$:
\begin{align}
    C(t)
    &= 
    \frac{(gm)^2f_\VS(t)}{8\sqrt{\pi} \sqrt{t(t-4m^2)}}
    \frac{\Gamma(1+\alpha_\tree (t))}
    {\Gamma(3/2+\alpha_\tree (t))}
    \no\\
    &\quad\times
    \lim_{s\to\infty}
    \left[
    \left(\frac{s_*^-}{s}\right)^{\alpha_\tree(t)}
    \sqrt{\frac{s^-_*}{s^+_*-s^-_*}}\,{}_2F_1\left(\frac{1}{2},\frac{1}{2};\alpha_\tree (t)+\frac{3}{2};\frac{-s^-_*}{s^+_*-s^-_*}\right)
    \right]\Theta(t-4m^2)
    \no\\
    &=
    \frac{(gm)^2f_\VS(t)}{8\sqrt{\pi} \sqrt{t(t-4m^2)}}
    \frac{\Gamma(1+\alpha_\tree (t))}
    {\Gamma(3/2+\alpha_\tree (t))}
    \no\\
    &\quad\times
    \left(\frac{\sqrt{\bar z}}{2}\right)^{\alpha_\tree(t)+1}
    g(\bar z,\alpha_\tree)\,
    {}_2F_1\left(\frac{1}{2},\frac{1}{2};\alpha_\text{tree}(t)+\frac{3}{2};\frac{1}{2}-\frac{1}{2\sqrt{\bar z-1}}\right)\Theta(t-4m^2)
    \,,\label{C_step1}
\end{align}
where $\bar z$ and $g(z,\alpha)$ are defined by 
\begin{align}
    \bar z \equiv \left(1+\frac{2{m'}^2}{t-4m^2}\right)^{-2}
    \,,\qquad
    g(z,\alpha)
    \equiv
    \frac{2^{\alpha+\frac{1}{2}}\left(1+\sqrt{1-z}\right)^{-\alpha}}
    {\sqrt{1-z+\sqrt{1-z}}}
    \quad
    (0<z<1)
    \,.
\end{align}
Since we are interested in the domain $t>4m^2$ and $m'>0$, it is sufficient to focus on the case $0<\bar z<1$. 
Next, we would like to express \eqref{C_step1} in terms of Legendre-Q function $Q_J(\bar z^{-1/2})$ as in \eqref{Cdef_general}. Let us recall that $Q_J$ is written in terms of the hypergeometric function ${}_2F_1$ as
\begin{align}
    Q_J(\bar z^{-1/2})=
    \frac{\sqrt{\pi } \Gamma (J+1) \, _2F_1\left(\frac{J}{2}+\frac{1}{2},\frac{J}{2}+1;J+\frac{3}{2};\bar z\right)}{(2z)^{J+1} \Gamma \left(J+\frac{3}{2}\right)}
    \,.\label{LQ_2f1}
\end{align}
For our purpose, it is convenient to establish the relation between the hypergeometric function in \eqref{LQ_2f1} and the one in \eqref{C_step1}.
We use the following two identities
\begin{align}
    \, _2F_1\left(a,a+\frac{1}{2};c;z\right)
    &=
    (1-z)^{-a}
    \, _2F_1\left(2 a,-2 a+2 c-1;c;\frac{1}{2}-\frac{1}{2\sqrt{1-z}}\right)
    \,,\label{2f1_identity1}
    \\
    \, _2F_1(a,b;c;z)
    &=
    (1-z)^{-a-b+c} \, _2F_1(c-a,c-b;c;z)
    \,.\label{2f1_identity2}
\end{align}
First, we set $a=(\alpha+1)/2$ and $c=\alpha+(3/2)$ in \eqref{2f1_identity1} to obtain 
\begin{align}
    _2F_1\left(\frac{\alpha+1}{2},\frac{\alpha}{2}+1;\alpha+\frac{3}{2},\bar z\right)
    =
    (1-\bar z)^{-\frac{\alpha+1}{2}}
    {}_2F_1\left(\alpha+1,\alpha+1;\frac{3}{2}+\alpha;\frac{1}{2}-\frac{1}{2\sqrt{\bar z-1}}\right)
    \,.\label{step1}
\end{align}
Next, we rewrite the RHS of \eqref{step1} by using \eqref{2f1_identity2} with $a=b=\alpha+1$ and $c=\alpha+(3/2)$. Then eq.~\eqref{step1} becomes
\begin{align}
    _2F_1\left(\frac{\alpha+1}{2},\frac{\alpha}{2}+1;\alpha+\frac{3}{2},\bar z\right)
    =
    g(\bar z,\alpha)\,
    {}_2F_1\left(\frac{1}{2},\frac{1}{2};\alpha+\frac{3}{2};\frac{1}{2}-\frac{1}{2\sqrt{\bar z-1}}\right)
    \quad
    (0<\bar z<1)
    \,.
    \label{eq:goal}
\end{align}
From eqs.~\eqref{eq:goal} and \eqref{LQ_2f1}, we find that \eqref{C_step1} equals to \eqref{Cdef_general} when $0<\bar z<1$.

\section{Finite energy sum rules for the Regge residue}\label{sec:FESR}
In this section, we explain how to obtain \eqref{eq:FESR0}. For convenience, we write our Regge amplitude in the $s\leftrightarrow u$ symmetric manner as
\begin{align}
    \disc_s\scat
    \approx 
    f(t) 
    \left(\frac{s-u}{8}\right)^{\alpha(t)}
    =
    f(t) 
    \left(\frac{s-2m^2+t/2}{4}\right)^{\alpha(t)}
    \,,
\end{align}
which is consistent at the leading order with the convention used in the main text in this paper $\disc_s\scat(s,t)\approx f(t)(s/4)^{\alpha(t)}$. The FESR for $f'$ derived in \cite{Noumi:2022wwf} is given by\footnote{Strictly speaking, the FESR for the scattering of identical massless scalar field is derived in \cite{Noumi:2022wwf}, though its extension to the case for massive scalar scattering is straightforward.}
\begin{align}
    \der_t \tilde f(t)|_{t=0}
    &
    \approx
    \int^{\Mstar^2}_4\frac{\mathrm{d}s}{s}\,
    x^2(18x^2-8)\partial_t\disc_{s}\scat(s,t)|_{t=0}
    \no\\
    &\quad 
    + \int^{\Mstar^2}_4\frac{\mathrm{d}s}{s}\,
    \frac{(-36x^3+27x^2+8x-4)x\disc_{s}\scat(s,0)}{\Mstar^2}
    \,,
    \quad
    x\equiv \frac{s-2m^2}{\Mstar^2}
    \,,\label{eq:tderivFESR1}
\end{align}
where $\tilde f(t)$ is the parameterization of Regge amplitude used in \cite{Noumi:2022wwf}, which is related to our $f(t)$ as 
\begin{align}
    \tilde f(t)
    =
    f(t)\left(\frac{\Mstar^2-2m^2+t/2}{4}\right)^{\alpha(t)}
    \,.\label{ftilde_def}
\end{align}
Let us now focus on the $m^{-2}$-enhanced term in \eqref{eq:tderivFESR1}. On the RHS of \eqref{eq:tderivFESR1}, the $m^{-2}$-enhanced term can come only from $\der_t\disc_s\scat(s,t)|_{t=0}$ in the first term as long as $\disc_s\scat(s,0)$ is not enhanced by $1/m$.  
We can also approximate $x$ as $x\simeq s/\Mstar^2$ in the integrand in \eqref{eq:tderivFESR1}. On the LHS of \eqref{eq:tderivFESR1}, we obtain the $m^{-2}$-enhanced term of $\der_t\tilde f$ only from the term proportional to $\der_tf$ after substituting \eqref{ftilde_def} into the LHS of \eqref{eq:tderivFESR1}. Thus, we can simplify \eqref{eq:tderivFESR1} into the following form when we focus on the $m^{-2}$-enhanced term
\begin{align}
    \der_tf_\oneloop^A(t)|_{t=0}
    \approx
    \left(\frac{4}{\Mstar^2}\right)^2
     \int^{\Mstar^2}_4\frac{\mathrm{d}s}{s}\,
    y^2(18y^2-8)\partial_t\disc_{s}\scat_A(s,t)|_{t=0}
    \,,\quad
    y\equiv s/\Mstar^2
    \,.\label{eq:FESR0pre}
\end{align}
Here, we replace $f$ and $\scat$ by $f_\oneloop^A$ and $\scat_A$, respectively, since at one-loop order only the type-A process can give the $m^{-2}$-enhanced term. Moreover, the $m^{-2}$-enhanced term of $\der_t\disc_s\scat_A$ can arise only from the first term of the dispersive integral \eqref{eq:discs_disp} for $\disc_s\scat_A$. Using this to rewrite $\der_t\disc_s\scat_A$ in \eqref{eq:FESR0pre} in terms of dispersive integral, we obtain \eqref{eq:FESR0}. 

Let us now explain how to evaluate the RHS of \eqref{eq:FESR0}. First, we substitute the exact expression \eqref{eq:jsum} for $\rho_{st}^A(s,t)$ into the integral over $t$ and obtain
\begin{align}
    \int^{t_*}_{4m^2}\frac{\mathrm{d}t}{\pi}
    \frac{\rho_{st}^A(s,t)}{t^2}
    &\simeq
    \frac{g^2}{\Mpl^2m^2}
     \int^{\frac{t_*}{4m^2}-1}_{0}\frac{\mathrm{d}\sigma_t}{\pi(\sigma_t+1)^2}
    \frac{1}{8\sqrt{\sigma_t (\sigma_t+1)}}
    \no\\
    &\quad\times
    \sum_{n=1}^\infty
    \frac{n^2 
    \Theta \left(s_*^-(s,t)-4n\right)}{\sqrt{\left(s_*^+(s,t)-4n\right)\left(s_*^-(s,t)-4n\right)}}
    \,,\label{integrand1}
\end{align}
where $\sigma_t\equiv(4m^2)^{-1}(t-4m^2)$ and the $m$-dependence of the residues of heavy particles is dropped since it is negligible. Then we substitute \eqref{integrand1} into the RHS of \eqref{eq:FESR0} to get
\begin{align}
    \text{[RHS of \eqref{eq:FESR0}]}
    \simeq
    &\frac{g^2}{\Mpl^2m^2}
    \left(\frac{4}{\Mstar^2}\right)^{2}
    \int^{\Mstar^2}_4\frac{\mathrm{d}s}{s}\,
    y^2(18y^2-8)
    \int^{\frac{t_*}{4m^2}-1}_{0}\frac{\mathrm{d}\sigma_t}{\pi(\sigma_t+1)^2}
    \frac{1}{8\sqrt{\sigma_t (\sigma_t+1)}}
    \no\\
    &\quad\times
    \sum_{n=1}^\infty
    \frac{n^2 
    \Theta \left(s_*^-(s,t)-4n\right)}{\sqrt{\left(s_*^+(s,t)-4n\right)\left(s_*^-(s,t)-4n\right)}}
    \,,\qquad
    y=s/\Mstar^2
    \,.\label{FESR_RHS}
\end{align}
From now we set $m'=m$. For $s>4$, we can use the approximation 
\begin{align}
    s_*^\pm(s,t)
    =s \left(2\sigma_t+1\pm\sqrt{4 \sigma_t+1}\right)/2\sigma_t+\mathcal{O}(s^0)
    \,.\label{starapp}
\end{align}
Under this approximation, the integrand of \eqref{FESR_RHS} becomes a function of $y $ and $\sigma_t$, which we can evaluate independently of $m$. We set an upper end of the $\sigma_t$-integral to be $1000$, though its precise value does not change the result as long as it is much greater than unity. This is because the $\sigma_t$-integral is dominanted by the contributions in the domain $\sigma_t\sim \mathcal{O}(1)$. In this way we can numerically evaluate the RHS of \eqref{eq:FESR0} and obtain the plot shown in fig.~\ref{fig:FESRtest_tderiv1}.

\section{Solving the \texorpdfstring{$t$}{t}-channel partial wave unitarity equation}\label{sec:reggetheory} 
In this section, we solve the unitarity equation \eqref{eq:unitarity} (or equivalently \eqref{unitarity3}) with
\begin{align}
    \phi_\ell(t)
    =
    \phi_\ell^\text{string}(t) + \phi_\ell^\text{eft}(t)
    \,,
\end{align}
where $\phi_\ell^\text{string}$ and $\phi_\ell^\text{eft}$ are given in \eqref{reg}. $\phi_\ell^\text{string}$ is a Regge pole at $\ell=\alpha(t)$ while $\phi_\ell^\text{eft}$ denotes a regular piece around $\ell=\alpha(t)$. For convenience, we introduce a function $\rho(t)$ which denotes the residue of $\phi_\ell^\text{string}$:
\begin{align}
    \phi_\ell^\text{string}(t)
    =
    \frac{\rho(t)}{\ell-\alpha(t)}
    \,,\qquad
    \rho(t)\equiv f(t)/r(\alpha(t))
    \,.\label{def_r}
\end{align}
Though our choice \eqref{reg} is specific to our setup, the analysis below and the result \eqref{eq:solution} is generic.

As already observed in section~\ref{sec:unitarity}, we should have $\disc_t\alpha\neq 0$ to prohibit the presence of double pole term on the RHS of \eqref{eq:unitarity}. Then, we have two limits $\ell\to \alpha(t_\pm)$ in which the both sides of \eqref{eq:unitarity} become singular. The singular term on the LHS has to be canceled with the singular term on the RHS in these limits. This requires
\begin{subequations}
    \label{eq:unitarity2}
\begin{align}
    &\frac{\rho(t_+)}{2i}
    =
    \left.
    \Phi_\ell(t)
    \left[
        \frac{\rho(t_+)\rho(t_-)}{\ell-\alpha(t_-)}
        + \rho(t_+)\phi_\ell^\text{eft}(t_-)
    \right]
    \right|_{\ell=\alpha(t_+)}
    \,,
    \\
    &\frac{-\rho(t_-)}{2i}
    =
    \left.
    \Phi_\ell(t)
    \left[
        \frac{\rho(t_+)\rho(t_-)}{\ell-\alpha(t_+)}
        + \rho(t_-)\phi_\ell^\text{eft}(t_+)
    \right]
    \right|_{\ell=\alpha(t_-)}
    \,.
\end{align}
\end{subequations}
The first and the second terms of \eqref{eq:unitarity2} are from the term $(\phi_\ell^\text{string})^2$ and the cross term $\phi_\ell^\text{string}\phi_\ell^\text{eft}$, respectively. Hence, each of them correspond to the B-term and the A-term defined in subsection~\ref{sec:unitarity}.  
It is straightforward to solve \eqref{eq:unitarity2} in terms of $\rho(t_\pm)$ as 
\begin{subequations}
\label{eq:solution}
\begin{align}
    &\disc_t \rho(t) 
    =
    \left[
        -\disc_t\left(\Phi^{-1}_{\alpha(t)}(t)\right)
        + \left(\phi_{\alpha(t_-)}^\text{eft}(t_+)+\phi_{\alpha(t_+)}^\text{eft}(t_-)\right)
    \right]\disc_t\alpha(t)
    \,,\\
    &\frac{\rho(t_+)+\rho(t_-)}{2}  
    =
    \left[
        \frac{1}{2}\left(\Phi^{-1}_{\alpha(t_+)}(t)+\Phi^{-1}_{\alpha(t_-)}(t)\right)
        +i\left(\phi_{\alpha(t_-)}^\text{eft}(t_+)-\phi_{\alpha(t_+)}^\text{eft}(t_-)\right)
    \right]\disc_t\alpha(t)    
    \,.
\end{align}
\end{subequations}

Our results \eqref{eq:solution} are generic. Let us now apply them to our specific setup to evaluate perturbative corrections to $(f,\alpha)$ for given leading-order terms $f_\VS$, $\alpha_\tree$, and $\phi_\ell^\tree$. Using the fact that $\alpha_\tree$ and the leading-order piece of $\phi_\ell^\text{eft}|_{\ell\in\mathbb{R}}$ are real at $t> 4m^2$, eqs.~\eqref{eq:solution} reduce to
\begin{subequations}
\label{eq:correction}
\begin{align}
    &\disc_t\, \rho(t) 
    =
    \left[
        \frac{\disc_t\alpha(t)}{\Phi_{\alpha_\tree}(t)}\ln\left(\frac{t-4m^2}{\Ms^2}\right)
        + 2\phi_{\alpha_\tree}^\text{eft}(t)
    \right]\disc_t\alpha(t)
    \,,\label{eq:correction_a}\\
    &\frac{\rho(t_+)+\rho(t_-)}{2} 
    =
        \frac{1}{\Phi_{\alpha_\tree}(t)}
    \disc_t\alpha(t)
    \,,\label{eq:correction_b}
\end{align}
\end{subequations}
at the leading order in coupling constants.
Thus, substituting the leading-order term of $\phi_\ell^\text{eft}$ in \eqref{reg} into eqs.~\eqref{eq:correction} and using \eqref{def_r} to rewrite the residue $\rho(t)$ in terms of $f(t)$, we obtain the following leading-order corrections:
\begin{subequations}
    \label{eq:result_reggetheory}
\begin{align}
    &\disc_tf(t)
    =
    f_\VS(t)\disc_t\alpha(t)\ln\left(\frac{t-4m^2}{\Ms^2}\right)
    + \frac{(gm)^2f_\VS(t)}{4\pi \sqrt{t(t-4m^2)}}
    Q_{\alpha_\tree(t)}\left(1+\frac{2{m'}^2}{t-4m^2}\right)
    \,,\label{eq:result_reggetheory_a}
    \\
    &\disc_t\alpha(t)
    =
    \frac{\Phi_{\alpha_\tree}(t)f_\VS(t)}{r(\alpha_\tree(t))}
    =
    \frac{f_\VS(t)}{2r(\alpha_\tree(t))}
    \sqrt{\frac{t-4m^2}{t}}
    \left(\frac{t-4m^2}{\Ms^2}\right)^{\alpha_\tree(t)}
    \,.\label{eq:result_reggetheory_b}
\end{align}
\end{subequations}
This solution includes corrections from both of the A term and the B term defined in \eqref{ABCterm}. The contributions from the B term can be extracted by setting $\phi_\ell^\text{eft}=0$ in \eqref{eq:result_reggetheory} as
\begin{align}
    &\disc_tf(t)|_\text{B term}
    =
    f_\VS(t)\ln\left(\frac{t{-}4m^2}{\Ms^2}\right)
    \disc_t\alpha(t)|_\text{B term}
    \,,\quad
    \disc_t\alpha(t)|_\text{B term}
    =
    \frac{\Phi_{\alpha_\tree}(t)f_\VS(t)}{r(\alpha_\tree(t))}
    \label{result_b}
    \,.
\end{align} 
This is identical to the solution of \eqref{unitarity3} with setting A term and C term to be zero. We then find that the leading-order results \eqref{eq:result_reggetheory} are simply given by the sum of A-term contributions and B-term contributions, 
\begin{subequations}
    \label{simple_sum}
\begin{align}
    &\disc_tf(t)
    =
    \disc_tf(t)|_\text{A term} + \disc_tf(t)|_\text{B term}
    \,,\\
    &\disc_t\alpha(t)
    =
    \disc_t\alpha(t)|_\text{A term} + \disc_t\alpha(t)|_\text{B term}
    \,,
\end{align}
\end{subequations}
where $\disc_tf(t)|_\text{A term}$ and $\disc_t\alpha(t)|_\text{A term}$ are given in \eqref{result_a} as a solution of \eqref{unitarity3} with setting B term and C term to be zero. Of course, a simple relation \eqref{simple_sum} is valid only at the leading-order in coupling constants, and the full solution \eqref{eq:solution} will not be simply written as the sum of contributions from each term in \eqref{unitarity3}.

\bibliography{posi2022.bib}

\end{document}